\documentclass[12pt]{article}
\usepackage{jheppub}
\pdfoutput=1
\usepackage{epstopdf}
\usepackage{mathrsfs}
\usepackage{psfrag}
\usepackage{epsf}
\usepackage{graphicx,epsfig}
\usepackage{amssymb}
\usepackage{epstopdf}
\usepackage{float}
\usepackage{braket}
\usepackage{hyperref}
\usepackage{xcolor}
\usepackage[backgroundcolor=white,linecolor=black]{todonotes}
\usepackage{amsmath,array,amsfonts,graphicx,wrapfig,lscape,float,multirow,longtable,rotating,subfigure,makecell}
\usepackage{url}

\newcommand{\be}{\begin{equation}}
\newcommand{\ee}{\end{equation}}
\newcommand{\beq}{\begin{equation}}
\newcommand{\beql}[1]{\begin{equation}\label{#1}}
\newcommand{\eeq}{\end{equation}}
\newcommand{\ba}{\begin{array}}
\newcommand{\ea}{\end{array}}
\newcommand{\bea}{\begin{eqnarray}}
\newcommand{\beal}[1]{\begin{eqnarray}\label{#1}}
\newcommand{\eea}{\end{eqnarray}}
\newcommand{\ben}{\begin{enumerate}}
\newcommand{\een}{\end{enumerate}}
\newcommand{\bean}{\begin{eqnarray*}}
\newcommand{\eean}{\end{eqnarray*}}

\newcommand{\btab}[1]{\begin{tabular}{#1}}
\newcommand{\etab}{\end{tabular}}

\newcommand{\comment}[1]{}

\newcommand{\qed}{\nobreak \ifvmode \relax \else
      \ifdim\lastskip<1.5em \hskip-\lastskip
      \hskip1.5em plus0em minus0.5em \fi \nobreak
      \vrule height0.75em width0.5em depth0.25em\fi}

\newcolumntype{C}[1]{>{\centering\arraybackslash}m{#1}}

\newcommand{\sign}{\textrm{sign}}

\def\beq{\begin{equation}}
\def\eeq{\end{equation}}

\newcolumntype{C}[1]{>{\centering\arraybackslash}m{#1}}
\newcommand{\N}{\mathcal{N}}

\newcommand{\comments}[1]{}

\newcommand{\pp}{\mathbf{p}}
\newcommand{\bx}{\mathbf{x}}

\def\Li{{\rm Li}}

\def\cM{{\cal M}}  
  
\def\cS{{\cal S}}  
  
\def\cY{{\cal Y}}

\renewcommand{\b}[1]{\braket{#1}}

\def\eps{\epsilon}

\graphicspath{{./images/}}

\def\beq{\begin{equation}}
\def\eeq{\end{equation}}
\def\bsp#1\esp{\begin{split}#1\end{split}}
\def\cM{{\cal M}}

\title{All two-loop MHV remainder functions in multi-Regge kinematics}

\author[a]{Vittorio Del Duca\footnote{On leave from INFN, Laboratori Nazionali di Frascati, Italy.}}
\author[b,c]{Claude Duhr}
\author[d]{Falko Dulat}
\author[b]{Brenda Penante}

\affiliation[a]{Institute for Theoretical Physics, ETH Z\"{u}rich, 8093 Z\"{u}rich, Switzerland.}
\affiliation[b]{Theoretical Physics Department, CERN, CH-1211 Geneva 23, Switzerland.}
\affiliation[c]{Center for Cosmology, Particle Physics and Phenomenology (CP3),\\
UCLouvain, B-1348, Louvain-La-Neuve, Belgium.}
\affiliation[d]{SLAC National Accelerator Laboratory,\\
Stanford University, Stanford, CA 94309, USA.}

\emailAdd{delducav@itp.phys.ethz.ch, c.duhr@cern.ch, dulatf@slac.stanford.edu, b.penante@cern.ch}

\abstract{
We introduce a method to extract the symbol of the coefficient of $(2\pi i)^2$ of MHV remainder functions in planar $\mathcal{N}=4$ Super Yang-Mills in multi-Regge kinematics region directly from the symbol in full kinematics. At two loops this symbol can be uplifted to the full function in a unique way, without any beyond-the-symbol ambiguities. We can therefore determine all two-loop MHV amplitudes at function level in all kinematic
    regions with different energy signs in multi-Regge kinematics. We analyse our results and we observe that they are consistent with the hypothesis of a 
    contribution from the exchange of a three-Reggeon composite state starting from two loops and eight points in certain kinematic regions.}

\preprint{
	\begin{flushright}CERN-TH-2018-256\\
	CP3-18-66\\
	SLAC-PUB-17357\end{flushright}\vspace{-0.4cm}
}

\begin{document}

\maketitle

\section{Introduction}
Tree-level scattering amplitudes in planar $\N=4$ Super Yang-Mills (SYM) are invariant under the action of a ``hidden'' dual superconformal symmetry in addition to the usual superconformal algebra~\cite{Drummond:2006rz,Drummond:2007au}. At loop level, the dual symmetry is anomalous due to the presence of infrared singularities and is preserved only by appropriate infrared safe quantities. One example is the \emph{remainder function}---the $L$-loop maximally helicity violating (MHV) amplitude subtracted of its infrared-divergent piece, in turn captured by the BDS-ansatz~\cite{Bern:2005iz}. Since the remainder is infrared finite, dual superconformal symmetry is manifestly unbroken and it is therefore a function of only the dual conformal cross ratios.

The remainder function has been widely studied using several complementary approaches. At the heart of many of these results lies the
fact that scattering amplitudes in planar $\mathcal{N}=4$ SYM are dual to lightlike polygonal
Wilson
loops~\cite{Alday:2007hr,Drummond:2007aua,Brandhuber:2007yx,Drummond:2007cf,Drummond:2007au,Drummond:2008aq,Alday:2007he,Berkovits:2008ic,CaronHuot:2010ek,Mason:2010yk}, which can in turn be computed as an operator product expansion (OPE) near the
collinear limit~\cite{Alday:2010ku,Gaiotto:2010fk,Gaiotto:2011dt,Sever:2011da}. The collinear OPE
was shown to be dual to the exchange of excitations of a flux tube sourced by the lightlike
polygonal frame of the Wilson loop. Through integrability methods, the spectrum of excitations as
well as their S-matrix can be determined for arbitrary values of the
coupling constant~\cite{Basso:2010in,Basso:2013vsa,Basso:2013aha,Basso:2014koa,Basso:2014jfa,Basso:2014nra,Basso:2014hfa,Basso:2015rta,Basso:2015uxa}. Furthermore, using the duality with Wilson loops the six-point MHV remainder function has been computed analytically at two loops~\cite{DelDuca:2009au,DelDuca:2010zg,Goncharov:2010jf}, and it has also enabled the determination of the symbols~\cite{Goncharov:2010jf,Chen:1977oja,GoncharovGrassmann,Brown:2009qja,Duhr:2011zq} and the total differential of all two-loop remainder functions for arbitrary multiplicity~\cite{CaronHuot:2011ky,Golden:2013lha}. At two loops fully analytic results with more than six points are only known for the seven-point MHV remainder function~\cite{Golden:2014xqf}. 

At higher loops the bootstrap program was used to constrain the six-point MHV and non-MHV amplitudes through six loops~\cite{Dixon:2011pw,Dixon:2011nj,Dixon:2013eka,Dixon:2014iba,Dixon:2014voa,Dixon:2015iva,Caron-Huot:2016owq}, as well as the symbol of the seven-point amplitude through three loops~\cite{Drummond:2014ffa,Dixon:2016nkn}. The main idea behind the bootstrap program is to make an educated guess for the function space in which the amplitude should live (e.g., by extrapolating from known lower loop results), and restrict the anatz for the amplitude by imposing general constraints coming from the known analytic structure of scattering amplitudes~\cite{Steinmann1,Steinmann2,Steinmann3,Gaiotto:2011dt,CaronHuot:2011ky}, from the conjecture that the symbol alphabet of scattering amplitudes in planar $\mathcal{N}=4$ SYM is related to the cluster algebra of certain Grassmannian spaces~\cite{Golden:2013xva,Golden:2014xqa,Golden:2014pua,Harrington:2015bdt,Drummond:2017ssj,Drummond:2018dfd,Golden:2018gtk} and the knowledge of the behaviour of amplitudes in certain kinematic limits. 
The study of amplitudes in particular kinematic limits is not only interesting as boundary data for the bootstrap program, but it is often the only way right now to gain analytic insight into the structure of scattering amplitudes with many loops and legs. In addition, these limits are often related to certain factorisation theorems, and the knowledge of the amplitudes in a limit may then allow one to extract the universal quantities (e.g., anomalous dimensions) that govern these factorisation theorems. 

The aim of this paper is to study scattering amplitudes in multi-Regge kinematics (MRK). This limit
describes scattering processes where the outgoing particles are strongly ordered in rapidity (or
equivalently along the lightcone).
In the Euclidean region, where all Mandelstam invariants are negative, scattering
amplitudes in MRK factorise to all orders in perturbation theory into certain building blocks. These
building blocks are described by resummed effective $t$-channel propagators (Reggeons) and the resummed
emission of strongly-ordered gluons along the ladder of effective propagators. They are determined
to all orders from four- and five-point amplitudes which are completely fixed by symmetry, and
therefore Euclidean scattering amplitudes in MRK are
trivial~\cite{Bartels:2008ce,Bartels:2008sc,Brower:2008nm,Brower:2008ia,DelDuca:2008jg}.
Starting from six points, scattering amplitudes in MRK are no longer trivial, if continued to
a particular kinematic region\footnote{That kinematic region is identified by a specific ordering of the energy signs of the outgoing particles.
We shall term a Mandelstam region any kinematic region with a generic ordering of the energy signs of the outgoing particles.}
before the limit is taken~\cite{Bartels:2008ce,Bartels:2008sc}.
The discontinuity is described by a dispersion relation similar to the BFKL equation in QCD~\cite{Kuraev:1976ge,Kuraev:1977fs,Balitsky:1978ic,Fadin:1998py,Camici:1997ij,Ciafaloni:1998gs}.
To leading logarithmic accuracy (LLA), scattering amplitudes in MRK can be described by an exchange of two Reggeons. More generally, the building blocks needed to describe the contributions to the amplitude from two-Reggeon exchange are known through next-to-leading logarithmic approximation (NLLA)~\cite{Bartels:2008sc,Lipatov:2010qg,Fadin:1998py,Bartels:2011ge,DelDuca:2018hrv}, and in the six-point case even to all orders in the 't Hooft coupling~\cite{Basso:2014pla}. Moreover, in refs.~\cite{Dixon:2012yy,DelDuca:2016lad} it was argued that to all loop orders for any number of points amplitudes in MRK in planar $\mathcal{N}=4$ SYM can be expressed in terms of single-valued
multiple polylogarithms~\cite{BrownSVHPLs,BrownSVMPLs,Brown:2013gia,Brown_Notes,DelDuca:2016lad}, in agreement with previously known analytic results~\cite{Bartels:2011ge,Lipatov:2010qg,Dixon:2011nj,Dixon:2011pw}. 

In the particular Mandelstam region where all produced partons but the two most forward ones have negative energy the amplitude is determined entirely by the exchange of composite states of two Reggeons even beyond LLA. In particular, we know analytic results for all seven- and eight-point three-loop amplitudes at NLLA~\cite{DelDuca:2018hrv,BramRobin}. In other Mandelstam regions and for more than seven particles, however, the amplitude is expected to receive contributions from the exchange of a composite state of three Reggeons. The building blocks describing these states are much less understood, and consequently not much is known about amplitudes in these regions. In ref.~\cite{Bargheer:2015djt} the symbols of all two-loop MHV amplitudes in any Mandelstam region where the two most forward particles have positive energy have been determined. It was however not possible to lift the symbols to function level, because the unknown three-Reggeon contribution is expected to be proportional to $(2\pi i)^2$, and so the symbol is (at least naively at first thinking) blind to three-Reggeon exchange. The corresponding three-loop symbols were analysed up to seven points in ref.~\cite{Bargheer:2016eyp}. 

In this paper we study two-loop amplitudes in any Mandelstam region, including all beyond-the-symbol terms. At the heart of our construction is the realisation that the symbol of an MHV remainder function is enough to determine the symbol of terms multiplied by $(2\pi i)^2$ in all Mandelstam regions. At two loops we can combine this result with the single-valuedness of remainder functions in MRK and lift the symbol to a unique function.
We obtain in this way complete analytic results for all two-loop MHV remainder functions in MRK in all Mandelstam regions. This complements the results of refs.~\cite{Bargheer:2015djt,DelDuca:2018hrv} where only the coefficients multiplying a single power of $2\pi i$ have been determined for a subset of Mandelstam regions. We find that the coefficients of $(2\pi i)^2$ have a very compact form and can be represented in terms of a simple combinatorial formula. By analysing explicit results through nine points, we observe that our results are consistent with the assumption of a contribution of a three-Reggeon composite state in certain Mandelstam regions, in agreement with the expectation from Regge theory.

The paper is organised as follows: In Section~\ref{sec:prelim} we present the kinematic setup of our
analysis and establish our notation for the multi-Regge limit. 
In Section~\ref{sec:symbol} we discuss the general structure of the remainder function in the
Euclidean region.
In Section~\ref{sec:regions} we parametrise the distinct Mandelstam regions and analyse the phase structure of the dual conformally-invariant cross ratios. We present a method to compute the symbol of the coefficient of $(2\pi i)^2$ in any given Mandelstam region from the symbol of the remainder function at any loop order.
In Section~\ref{sec:results} we focus on two loops and present explicit results obtained from analytically continuing the symbol of the
two-loop remainder function for a high number of points. We discuss our findings for the structure
and relations of our results of different Mandelstam regions.
In Section~\ref{sec:discussion} we analyse our explicit two-loop results through nine points and we  argue that our results are consistent with the assumption of a contribution of a three-Reggeon composite state in certain Mandelstam regions.
Finally, we summarise our findings in Section~\ref{sec:conclusions}.

\section{Preliminaries}
\label{sec:prelim}
In this paper we consider scattering amplitudes in planar $\mathcal{N}=4$ Super Yang-Mills (SYM) in multi-Regge kinematics (MRK), i.e., $2 \rightarrow N-2 $ scattering processes where the outgoing particles are strongly ordered in rapidity. If we parametrise the momenta in terms of lightcone and transverse coordinates, as defined in Appendix~\ref{app:momentum-twistor-MRK},
\beq
p_i \equiv (p_i^+,p_i^-,\pp_i)\ ,\quad 1\leq i \leq N\ ,
\eeq
then in MRK the momenta of the outgoing particles $3,\ldots,N$ are ordered as follows,
\beq
\label{eq:MRK}
|p_3^+|\gg \ldots\gg |p_N^+|,\quad 	|p_3^-|\ll \ldots\ll |p_N^-|\quad \text{and}\quad |\pp_{3}|\simeq\ldots\simeq |{\pp}_{N}|\ .
\eeq

The goal of this paper is to compute all MHV remainder functions at two loops in MRK. In order to achieve this, we start from the study of their symbols in general kinematics given in ref.~\cite{CaronHuot:2011ky}. Dual conformal symmetry implies that the symbols are functions of dual conformal cross ratios,
\begin{align}
\label{eq:Us}
U_{ijkl} =\frac{x_{ij}^2 x_{kl}^2}{x_{ik}^2 x_{jl}^2}\ ,\quad U_{ij}\equiv U_{i \, j+1 \, j\, i+1}\ ,
\end{align}
where $x_i$ are dual variables which parametrise a set of $N$ particles satisfying momentum conservation,
\begin{align}
\label{eq:x}
p_i = x_i - x_{i-1}\,, \quad x_{ij}\equiv x_i-x_j\ .
\end{align}

Due to the strong ordering given in eq.~\eqref{eq:MRK}, in MRK the non-trivial dynamics takes place in the transverse space with respect to the two incoming high-energy particles. It is therefore useful to consider transverse dual variables for the $N-2$ outgoing particles,
\beq
\pp_i = \bx_{i-1}-\bx_{i-2}\,,\quad \bx_{N-1} \simeq \bx_{1} \,,\quad i = 3,\dots N \ .
\eeq

\begin{figure}[h]
	\centering
	\includegraphics[width=0.6\linewidth]{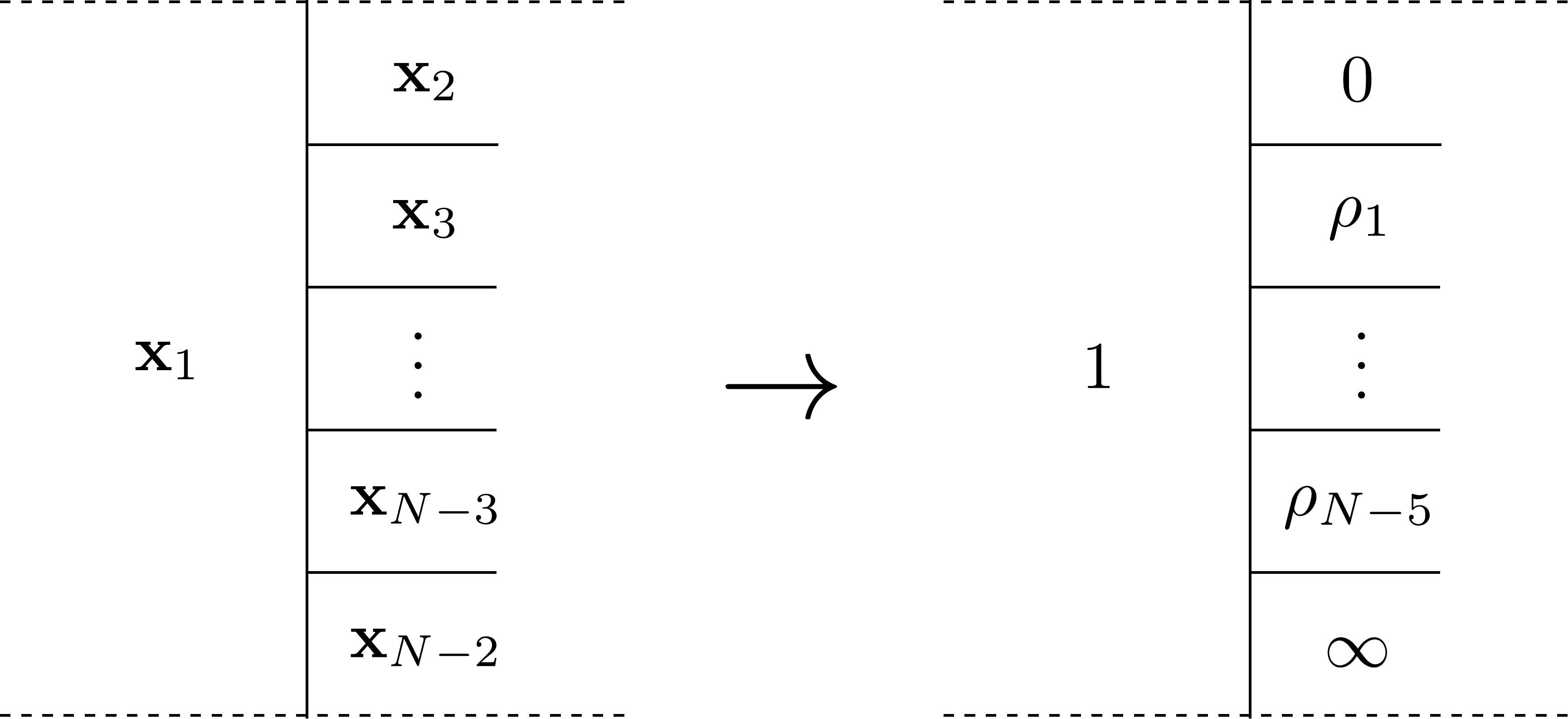}
	\caption{Simplicial MRK coordinates parametrising the transverse space for the scattering of $N$ particles in MRK.}
	\label{fig:rho}
\end{figure}

The space of kinematic configurations in the two-dimensional transverse space of a scattering of $N$ particles in MRK is captured by the moduli space $\mathfrak{M}_{0,N-2}$ of Riemann spheres with $N-2$ marked points~\cite{DelDuca:2016lad}. Dual conformal transformations act on the transverse space as M\"obius transformation for $SL(2,\mathbb{C})$. We can therefore choose a local system of coordinates where three of the $\bx_i$ variables are set to 0, 1 and $\infty$. A convenient choice which we will adopt throughout is $\{\bx_1,\bx_2,\dots,\bx_{N-2} \}= \{1,0,\rho_1,\dots,\rho_{N-5},\infty\}$, shown in fig.~\ref{fig:rho} and referred to as \emph{simplicial MRK coordinates} in ref. \cite{DelDuca:2016lad}.

In the following it will be useful to classify the cross ratios $U_{ij}$ into three non-overlapping classes, 
\begin{align}
\bsp 
\label{eq:crMRK}
	V_i\,&\equiv\,U_{1i}\,,\,\,\qquad 4\le i\le N-2\, , \\
	\tilde{V}_j\,&\equiv\,U_{jN}\,,\qquad 3\le j\le N-3\, ,\\
	W_{ij}\,&\equiv\,U_{ij}\,, \qquad\;\; i,j\notin\{1,N\}\,.
	\esp
	\end{align}
If we introduce an operator $\cM$ corresponding to taking the multi-Regge limit of an expression,
\beq
\cM\left[X\right]\,\equiv \lim_{\textrm{MRK}} X\,,
\eeq
then the multi-Regge limit of the cross ratios above takes the form
\beq
\label{eq:crMRKprime}
	\cM\left[V_i\right]=\,\cM[\tilde{V}_j]\,= 0\,,\quad \cM\left[W_{ij}\right]\,= \,1\ .
\eeq
The cross ratios $U_{ij}$ form a spanning set of all possible cross ratios $U_{ijkl}$ that we can form out of $N$ points\footnote{The relations between the cross ratios $U_{ijkl}$ and $U_{ij}$ can be found in Appendix \ref{app:momentum-twistor-MRK}.}, therefore we can derive the multi-Regge limit of a generic cross ratio from eq.~\eqref{eq:crMRKprime}. We find
\begin{align}
\bsp
\label{eq:crMRK2}
\cM[U_{1kjl}]\, &\,= 0 \,,\quad 1<j<k,\,1<l\\
\cM[U_{ikjl}]\, &\,= 1-|f_{ijkl}(\rho)|^2 \prod_m \tau_m\, ,\quad 1<i<j<k \textrm{ and } i<l\ 
,
\esp
\end{align}
where $f_{ijkl}(\rho)$ is a function of the simplicial MRK coordinates $\rho_i$ whose form depends on the
cross ratio under consideration. The rate at which the cross ratios approach 1 is parametrised by
\beq
\label{eq:tau}
\tau_i\equiv\pm\sqrt{V_{i+3}\tilde{V}_{i+2}}\,, \quad 1\leq i \leq N-5\ .
\eeq
As such, the direction from which they approach 1 depends on the signs of the $\tau_i$ variables, which in turn depend on the energies of the outgoing particles as follows,
\beq
\label{eq:tausign}
\sign(\tau_i)\,=\,\sign\left(\frac{p_{i+4}^+}{p_{i+3}^+}\right)\,=\,\varrho_{i+3} \,\varrho_{i+4}\ ,
\eeq
where 
\beq
\label{eq:varrho}
\varrho_i\equiv \sign(p_i^0) = \sign(E_i)\,, \qquad 3\le i\le N\,.
\eeq



\section{Remainder functions in the Euclidean region}
\label{sec:symbol}

In this section we study the symbols of MHV remainder functions in MRK in the Euclidean region where all consecutive Mandelstam variables are negative. The results of this section are in principle well known. In particular it is well known that the remainder function vanishes in MRK in the Euclidean region where all consecutive Mandelstam invariants are negative~\cite{Bartels:2008ce,Bartels:2008sc,Brower:2008nm,Brower:2008ia,DelDuca:2008jg}. We recall here in detail the analytic structure of remainder functions in MRK, because this is the foundation to correctly perform the analytic continuation of remainder functions to other Mandelstam regions.

If we take into account the first and second entry conditions
\cite{Gaiotto:2011dt,CaronHuot:2011ky}, we can write the symbol of the $L$-loop $N$-point remainder
function $R_N^{(L)}$ without loss of generality in the form
\beq\bsp\label{eq:RNL_symbol}
\cS\left(R_N^{(L)}\right)&\, = \sum_{\substack{i<j<l}}(U_{ij}\otimes U_{il}+U_{il}\otimes U_{ij})\otimes A_{ijil}+\sum_{\substack{i<j\\i< k<l}}(U_{ij}\otimes U_{kl}+U_{kl}\otimes U_{ij})\otimes A_{ijkl}\\
&\, + \sum_{\substack{i<j<k\\i<l}}U_{ikjl}\otimes (1-U_{ikjl})\otimes B_{ijkl}\,,
\esp\eeq
where $A_{ijkl}$ and $B_{ijkl}$ denote integrable tensors of weight $2(L-1)$ that involve more complicated letters.
Note that with the chosen summation ranges this representation is unique.
Equivalently, we may write the previous equation as a statement about the $(2,1,1,\ldots,1)$ component of the coaction on the remainder function,
\beq\bsp\label{eq:Eucl211}
\Delta_{2,1,\ldots,1}\left(R_N^{(L)}\right)&\, = \sum_{\substack{i<j<l}}(\log U_{ij}\,\log U_{il})\otimes A_{ijil}+\sum_{\substack{i<j\\i<k<l}}(\log U_{ij}\,\log U_{kl})\otimes A_{ijkl}\\
&\, - \sum_{\substack{i<j<k\\i<l}}\textrm{Li}_2(1-U_{ikjl})\otimes B_{ijkl} + \zeta_2\otimes C\,,
\esp\eeq
where $C$ denotes an integrable tensor of weight $2(L-1)$ that is not determined from the knowledge of eq.~\eqref{eq:RNL_symbol} alone.\footnote{It is constrained though by requiring $R_N^{(L)}$ to have the correct behaviour in certain limits. Note that $C$ does not satisfy any obvious first or second entry conditions.} 

We now study the multi-Regge limit of eq.~\eqref{eq:Eucl211}. It is convenient to write eq.~\eqref{eq:Eucl211} explicitly in terms of the cross ratios $V_i,\,\tilde{V}_j$ and $W_{ij}$ defined in eq. \eqref{eq:crMRK},
\beq\bsp
\label{eq:211}
&\Delta_{2,1,\ldots,1}\left(R_N^{(L)}\right) = \sum_{\substack{i<j}}\left[(\log V_{i}\,\log V_{j})\otimes A_{1i1j}+(\log \tilde{V}_{i}\,\log \tilde{V}_{j})\otimes A_{iNjN}\right]\\
&\,
+\sum_{\substack{1<i,j<N}}(\log V_{i}\,\log \tilde{V}_{j})\otimes A_{1ijN} +\sum_{\substack{1<i\\1<j<k<N}}(\log V_{i}\,\log W_{jk})\otimes A_{1ijk}\\
&\,+\sum_{\substack{1<i<j\\1<i<k<N}}(\log W_{ij}\,\log \tilde{V}_{k})\otimes A_{ijkN} +\sum_{\substack{1<i<j<l<N}}(\log W_{ij}\,\log W_{il})\otimes A_{ijil}\\
&\,+\sum_{\substack{1<i<j<N\\i<k<l<N}}(\log W_{ij}\,\log W_{kl})\otimes A_{ijkl}- \sum_{\substack{1<j<k\\1<l}}\textrm{Li}_2(1-U_{1kjl})\otimes B_{1jkl} \\
&\,- \sum_{\substack{i<j<k\\1<i<l}}\textrm{Li}_2(1-U_{ikjl})\otimes B_{ijkl} + \zeta_2\otimes C\ .
\esp\eeq
Using eqs.~\eqref{eq:crMRKprime} and \eqref{eq:crMRK2} it is now straightforward to obtain the multi-Regge limit of eq.~\eqref{eq:211}. We see that all terms that contain a cross ratio $W_{ij}$ in the first entry of the coaction vanish, and we find
\beq\bsp
\label{eq:211MRL}
&\Delta_{2,1,\ldots,1}\left(\cM\left[R_N^{(L)}\right]\right) = \sum_{\substack{i<j}}\left[(\log V_{i}\,\log V_{j})\otimes \cM[A_{1i1j}]+(\log \tilde{V}_{i}\,\log \tilde{V}_{j})\otimes \cM[A_{iNjN}]\right]\\
&\,
+\sum_{\substack{1<i,j<N}}(\log V_{i}\,\log \tilde{V}_{j})\otimes \cM[A_{1ijN}]+\zeta_2\otimes\cM\Big[C - \sum_{\substack{1<j<k\\1<l}} B_{1jkl}\Big]\ .
\esp\eeq
Note the additional contribution to the $\zeta_2$ term due to the vanishing of $\cM[U_{1jkl}]$ in
the argument of the dilogarithm.
Since the remainder function vanishes in MRK in the Euclidean region, it follows that the coefficients above must satisfy
\beq\bsp\label{eq:M_ABC}
\cM[A_{1i1j}] = \cM[A_{1ijN}] = \cM[A_{iNjN}] =0 {\rm~~and~~} \cM[C] = \sum_{\substack{1<j<k\\1<l}} \cM[B_{1jkl}]\,.
\esp\eeq
Before we extend this discussion to other Mandelstam regions in the next section, let us make some comments about eq.~\eqref{eq:M_ABC}. 
First, we see that the value of $C$ in MRK is fixed from the symbol in MRK. Second, although eq.~\eqref{eq:M_ABC} was derived in the Eucliden region, these relations hold for symbols in \emph{any} Mandelstam region, because the symbol of a function is independent of branch cuts and the region in which the function is evaluated. These two observations combined have an important consequence, which will be the key to compute the coefficient of $(2\pi i)^2$ from the symbol in any Mandelstam region. In general kinematics, the coefficient of $\pi^2$ is a beyond-the-symbol term, which is reflected in the fact that the value of $C$ in eq.~\eqref{eq:Eucl211} is not determined by the symbol alone. In MRK, however, the value of $\cM[C]$ is uniquely determined, independently of the kinematic region! We can therefore access the (symbol of the) coefficient of $(2\pi i)^2$ in the remainder function without the explicit knowledge of the value of $C$ in general kinematics. In the next section we show how we can combine this observation with the correct analytic continuation of the remainder function to determine the symbol of the coefficient of $(2 \pi i)^2$ in MRK.


\section{Remainder functions in different Mandelstam regions}
\label{sec:regions}
\subsection{Cross ratios and Mandelstam regions}

We now discuss the analytic continuation to kinematic regions where particles 1 and 2 are incoming with
positive energy and the signs of the energies of the remaining particles are not fixed. Following
refs.~\cite{Bartels:2014mka,Bargheer:2015djt} we define a \emph{Mandelstam region} by specifying the signs of
the energies of particles $3$ to $N$, labeling them by the sequence $\varrho \equiv(\varrho_i)_{3\le i\le N}$ with $\varrho_i$ given in eq.~\eqref{eq:varrho}. 
The data defining a region does not yet uniquely fix the signs of the Mandelstam invariants, but only the signs of two-particle invariants are fixed.
Since $s_{ij} = E_i\,E_j(1-\cos\theta_{ij})$, we have
\beq
\sign(s_{ij}) = \sign(E_i)\,\sign(E_j) = \varrho_i\,\varrho_j\,.
\eeq
We thus see that the signs of two-particle invariants are uniquely specified by the signs of the energies of the produced particles, i.e., each two-particle invariant has a unique sign in a given Mandelstam region $\varrho$.
For multi-particle invariants, however, this is in general not the case. We now show that in MRK the sign of each consecutive invariant $x_{ij}^2$ is uniquely fixed by the signs of the energies of the external momenta.
We start by noting that in MRK we have
\beq
x_{ij}^2\,\prod_{l=i+2}^{j-1}\kappa_{l} = s_{(i+1)\ldots j}\,\prod_{l=i+2}^{j-1}\kappa_{l}\simeq \prod_{k=i+1}^{j-1} s_{k(k+1)}\,,\quad \forall\, 2\le i<j\le N\,,
\eeq
with 
\beq
\kappa_{j} = |{\bf p}_{j}|^2\,,\quad 4\le j\le N-1\,.
\eeq
Since $\kappa_{j}$ is always a positive number and $\varrho_i^2=1$, we have
\beq\label{eq:x_sign}
\sign(x_{ij}^2) = \prod_{k=i+1}^{j-1} \sign(s_{k(k+1)}) = \varrho_{i+1}\,\varrho_j\,.
\eeq
Note that eq.~\eqref{eq:x_sign} remains valid for two-particle invariants.
If in addition we require that all $t$-channel invariants are negative, $x_{1i}^2<0$, then we see that in MRK the signs of the multi-particle invariants are fixed. Therefore, in order to study the amplitude in MRK in the Mandelstam region $\varrho$, we should first analytically continue the amplitude to the kinematic region where all invariants have the signs prescribed by eq.~\eqref{eq:x_sign}. Then, from this kinematic region the multi-Regge limit can be reached without any invariants changing sign. Since we are only interested in MRK, we will from now on always assume that all invariants take the signs prescribed by eq.~\eqref{eq:x_sign} when talking about the corresponding region $\varrho$.

%

Let us now discuss what happens to the cross ratios when we analytically continue them from the Euclidean region to a Mandelstam region.
The analytic structure of an amplitude is such that each consecutive Mandelstam invariant has a small positive imaginary part, $x_{jk}^2 \to x_{jk}^2 + i\varepsilon$. Moreover, the amplitude is real in the Euclidean region where all consecutive Mandelstam invariants are negative and the cross ratios $U_{ijkl}$ have no phase in the Euclidean region. If the Mandelstam invariants change sign, however, the cross ratios may acquire a region-dependent phase that can be parametrised as
\beq
U_{ijkl} = \left|\frac{x_{ij}^2 x_{kl}^2}{x_{ik}^2 x_{jl}^2}\right|\,\exp\left[-i\pi (\varphi_{ij}+\varphi_{kl}-\varphi_{ik}-\varphi_{jl})\right]\,,
\eeq
with
\beq
\varphi_{ij} = \theta(x_{ij}^2) = \left\{\begin{array}{ll}
	1\,,& \textrm{ if } x_{ij}^2>0\,,\\
	0\,,& \textrm{ if } x_{ij}^2<0\,.
\end{array}\right.
\eeq

The phase structure of the cross ratios $U_{ij}$ in
a Mandelstam region $\varrho$ can be described in terms of winding numbers~\cite{Bargheer:2015djt}, defined by\footnote{Our winding numbers differ by an overall sign from ref.~\cite{Bargheer:2015djt}.}
\beq
U_{jk} \to |U_{jk}|\,\exp\left[2\pi i\,n^{\varrho}_{jk}\right]\,.
\eeq
The winding numbers $n_{jk}^\varrho$ for the cross ratios defined in eq.~\eqref{eq:crMRK} are determined as follows.
\begin{enumerate}
	\item For the cross ratios $V_i$ and $\tilde{V}_j$ we have
	\beq
	\label{eq:windingsmall}
	n_{1j}^{\varrho} = -n_{(j-1)N}^{\varrho} = -\frac{1}{4}\,(\varrho_j-\varrho_{j+1}) = \left\{\begin{array}{ll}
		\phantom{-}0\,,&\textrm{ if } \varrho_j = \varrho_{j+1}\,,\\
		-1/2\,,&\textrm{ if } \varrho_j = -\varrho_{j+1}=-1\,,\\
		\phantom{-}1/2\,,&\textrm{ if } \varrho_j = -\varrho_{j+1}=+1\,.
	\end{array}\right.
	\eeq
    The winding number of any cross ratio $V_i$ or $\tilde{V}_j$ is either vanishing or
    half-integer, which means that they can acquire at most a phase of $\pm i\pi$. Note that depending on the region they can change sign.
\item For the cross ratios $W_{jk}$ we have,
	\beq
	\label{eq:windinglarge}
	n_{jk}^{\varrho} = -\frac{1}{4}\,(\varrho_{j+1}-\varrho_{j+2})\,(\varrho_{k+1}-\varrho_{k})\,.
	\eeq
    The above winding numbers are always 0 or $\pm1$ and thus $W_{jk}$ cannot change sign.
\end{enumerate}
In a similar fashion, the winding numbers of the cross ratios $U_{ijkl}$ can be determined from the knowledge of those for
the $U_{ij}$ using the expressions found in Appendix~\ref{app:momentum-twistor-MRK}.


\subsection{Symbols of remainder functions in MRK}
	\label{sec:MRKotherregions}
	
	In this section we present one of the main results of this paper, namely a method to extract the symbol of the coefficient of $(2\pi i)^2$ of the $L$-loop $N$-point remainder function $R_N^{(L)}$ in MRK in any Mandelstam region just from the knowledge of the symbol of $R_N^{(L)}$.
	
	We have to analytically continue $R_N^{(L)}$ to the Mandelstam region $\varrho$ before taking the multi-Regge limit. After analytic continuation, the remainder function has the general form
	\begin{align}
	\label{eq:general-remainder}
	\begin{split}
	R^{(L)\varrho}_N &= \sum_{k=0}^{2L}\,(2 \pi i)^k T^{(L)\varrho}_{k,N}\, .
	\end{split}
	\end{align}
	We can then approach the multi-Regge limit, and we obtain
	\beq
	\cM\left[R^{(L)\varrho}_N\right] = \sum_{k=1}^{2L}\,(2 \pi i)^k X^{(L)\varrho}_{k,N}\,,\qquad \cM\left[T^{(L)\varrho}_{k,N}\right] = X^{(L)\varrho}_{k,N}\,.
	\eeq
	where the $X^{(L)\varrho}_{i,N}$ are weight $(2L-i)$ real and single-valued functions of the $3(N-5)$ variables $\{\rho_i,\bar{\rho}_i,\tau_i\}$, where $\rho_i$ are the dual variables shown in fig.~\ref{fig:rho} and $\tau_i$ are defined in eq.~\eqref{eq:tau}. We have dropped the contribution from $k=0$, because the remainder functions vanishes in MRK in the Euclidean region, and so all non-vanishing terms must be proportional to at least one power of $(2\pi i)$. They  are symmetric under target-projectile exchange \cite{Bartels:2012gq,Bartels:2014mka},
	\begin{align}
	\bsp 
	(\rho_i,\bar{\rho}_i,\tau_i)\mapsto(1/\rho_{N-4-i},1/\bar{\rho}_{N-4-i},\tau_{N-4-i})\,,\quad 1\leq i \leq N-5\ .
	\esp
	\end{align}
Since the single-valued version of $2\pi i$ is zero, the single-valuedness of $X^{(L)\varrho}_{i,N}$ implies that the coefficient of each power of $2\pi i$ is uniquely defined, and no additional powers of $\pi$ are generated when approaching a soft limit. As such, they independently satisfy soft limits,
	\begin{align}\label{eq:X_soft}
	\lim_{{\bf x}_{j}\rightarrow {\bf x}_{j+1}} X^{(L)\varrho}_{k,N+1}\, =  \, X^{(L)\varrho'}_{k,N}\,,\quad 1\leq k\leq 4 \,,\quad 2\le j\le N-3 \ ,
	\end{align}
	where $\varrho'$ is obtained from $\varrho$ by deleting $\varrho_{j+2}$.

	We now discuss how we can extract the symbols of $X^{(L)\varrho}_{1,N}$ and $X^{(L)\varrho}_{2,N}$ from the symbol of $R^{(L)}_N$. Performing the analytic continuation from the Euclidean region to the Mandelstam region $\varrho$ in general kinematics is an extremely complicated task, and it is in general not known how to perform this analytic continuation. Performing the analytic continuation of $\Delta_{2,1,\ldots,1}\left(R_N^{(L)}\right)$, however, is very simple. Indeed, analytic continuation only acts in the first entry of the coaction~\cite{Duhr:2012fh,Brown:coaction}. The first entries of $\Delta_{2,1,\ldots,1}\left(R_N^{(L)}\right)$ are very simple and contain only logarithms and dilogarithms of the form $\log U_{ij}$ and $\Li_2\left(1-U_{ijkl}\right)$ (cf.~eq.~\eqref{eq:Eucl211}), whose analytic continuations are well known~\cite{Duplancic:2000sk,vanHameren:2005ed}:
	
	\begin{align}
	\begin{split}
	\log U_{ij}\,=\,&\log\left|U_{ij}\right| + 2\pi i\, n^{\varrho}_{ij}\ ,\\[5pt]
	\label{eq:Li2ac}
	\Li_2\left(1-U_{ijkl}\right) \,=\,&  \text{Re}\,\text{Li}_2\left(1-U_{ijkl}\right) -2\pi i\, n_{ijkl}^{\varrho} \log\left|1-U_{ijkl}\right|\\
	&-\frac{1}{2}\,(2\pi i)^2 (n^\varrho_{ijkl})^2\,\theta\left(U_{ijkl}-1\right),
	\end{split}
	\end{align}
	where the winding numbers are defined in eqs.~\eqref{eq:windingsmall} and~\eqref{eq:windinglarge}. We can thus insert eq.~\eqref{eq:Li2ac} into eq.~\eqref{eq:Eucl211} and then take the multi-Regge limit. Note that the contribution from $C$ in eq.~\eqref{eq:Eucl211} drops out in MRK indepedently of the Mandelstam region once we enforce eq.~\eqref{eq:M_ABC}. The symbols of $X^{(L)\varrho}_{1,N}$ and $X^{(L)\varrho}_{2,N}$ can then easily be read off by collecting the terms proportional to $2\pi i$ and $(2\pi i)^2$ in the first entry of the coaction.
	
We see that we can determine the symbols of $X^{(L)\varrho}_{1,N}$ and $X^{(L)\varrho}_{2,N}$ from the knowledge of the symbol of $R_N^{(L)}$. Crucial in this procedure is the fact that analytic continuation only acts in the first factor of the coaction on MPLs. The analytic continuation is trivialised by the fact that the first and second entry conditions on the symbol of $R_N^{(L)}$ force the first factor in $\Delta_{2,1,\ldots,1}\left(R_N^{(L)}\right)$ to have a very simple structure. We emphasise that the analytic continuation is free of any ambiguity and does not rely on any choice of path: the analytic continuation is entirely fixed by the Feynman $+i\varepsilon$ prescription, as expected for any physical scattering amplitude.
	
	Let us conclude this section by commenting on the symbols of $X^{(L)\varrho}_{k,N}$ with $k>2$. In principle, the symbols of these functions could be extracted in a similar fashion from $\Delta_{k,1,\ldots,1}\left(R_N^{(L)}\right)$. In practise, however, it is known that starting from the third entry the symbol of $R_N^{(L)}$ involves letters that are not simply $U_{ij}$ or $1-U_{ijkl}$, but more general algebraic functions of the conformal cross ratios appear. As a consequence, the first factor in $\Delta_{k,1,\ldots,1}\left(R_N^{(L)}\right)$ will no longer involve only simple functions whose analytic continuation to arbitrary Mandelstam regions is known in the literature. For example, for $N=6$, it is known that the first factor in $\Delta_{3,1,\ldots,1}\left(R_6^{(L)}\right)$ involves the one-loop hexagon in six dimensions~\cite{Dixon:2013eka,Dixon:2011ng,DelDuca:2011ne}, and it is currently not known how to analytically continue this function to arbitrary Mandelstam regions.

	While in general we cannot constrain the symbols of $X^{(L)\varrho}_{k,N}$ with $k>2$, we can make concrete predictions for the cases $k=2L-1$ and $k=2L$. First, $X^\varrho_{2L,N}$ must have weight zero, and so it is a rational number. Equation~\eqref{eq:X_soft} then implies that $X^{(L)\varrho}_{2L,N}$ is independent of $N$, and it must be equal to $X^{(L)\varrho}_{2L,6}$ (it does depend, at least a priori, on the number of loops and the Mandelstam region). 
    
    The function $X^{(L)\varrho}_{2L-1,N}$ has weight one and is thus no longer forced to be constant. We now show that the constraints imposed by single-valuedness and eq.~\eqref{eq:X_soft} force $X^{(L)\varrho}_{2L-1,N}$ to vanish.
        $X^{(L)\varrho}_{2L-1,N}$ is a single-valued pure function of weight one on $\mathfrak{M}_{0,N-5}$, and as such can be expanded in a basis of single-valued logarithms,
	\begin{align}
	\bsp 
	&\log|\rho_i|^2,\quad \log|1-\rho_i|^2,\qquad 1\leq i \leq N-5\ .\\
	&\qquad\log\left|1-\frac{\rho_i}{\rho_j}\right|^2,\qquad\qquad 1\leq i<j\leq N-5\ .
	\esp
	\end{align}
    Thus, in general we can write,
	\begin{align}
	\bsp 
	\label{eq:X3N}
	X^{(L)\varrho}_{2L-1,N}\,&=\,\sum_{i=1}^{N-5} \left( \alpha_i \log|\rho_i|^2 + \beta_i \log|1-\rho_i|^2 + \sum_{j=i+1}^{N-5} \gamma_{ij} \log\left|1-\frac{\rho_i}{\rho_j}\right|^2 \right)\ ,
	\esp
	\end{align}
	for constants $\alpha_i,\,\beta_i$ and $\gamma_{ij}$ (their value may still depend on $N$, $L$ and $\varrho$). 
	We now show by induction that $X^{(L)\varrho}_{2L-1,N}$ must vanish for all $N$. The argument is very similar to the argument in the two-loop case in ref.~\cite{DelDuca:2018hrv}.
	The start of the recursion is $N=6$, where eq.~\eqref{eq:X3N} reduces to
	\begin{align}
	\bsp 
	X^{(L)\varrho}_{2L-1,6}\,&=\,  \alpha \log|\rho_1|^2 + \beta_1 \log|1-\rho_1|^2\ ,
	\esp
	\end{align}
	As the six-point remainder function must vanish in all soft limits, eq.~\eqref{eq:X_soft} implies
	\beq
	\lim_{\rho_1\to 0}X^{(L)\varrho}_{2L-1,6} = \lim_{\rho_1\to \infty}X^{(L)\varrho}_{2L-1,6} = 0\,.
	\eeq
	The only way to satisfy these constraints is that $X^{(L)\varrho}_{2L-1,6}$ vanishes identically.
	Next, assume that $X^{(L)\varrho}_{2L-1,N'}$ vanishes for $N'<N$. 
	We can apply different soft limits to the $N$-particle ansatz and derive constraints on the coefficients. Consider first the soft limit $\rho_1 \rightarrow 0$. Since $X^{(L)\varrho}_{2L-1,N-1}=0$ by induction, all terms in eq.~\eqref{eq:X3N} that do not vanish in this limit must appear accompanied by a vanishing coefficient, and so
	\begin{align}
	\bsp
	\alpha_i &= 0\,,\quad 1\leq i \leq N-5\ ,\\
	\beta_i &= 0 \,,\quad 2\leq i \leq N-5\ ,\\
	\gamma_{ij}&=0\,,\quad i\neq 1\ .
	\esp
	\end{align}
	Now, consider the soft limit $\rho_1\rightarrow \rho_2$. From this we derive the additional constraints
	\begin{align}
	\bsp 
	\beta_1 &= -\beta_2\quad \Rightarrow\quad \beta_1 = 0\ ,\\
	\gamma_{12} &= 0\,,\quad	\gamma_{1j} = -\gamma_{2j},\quad j>2 \quad \Rightarrow\quad \gamma_{1j} = 0\ .
	\esp
	\end{align}
	Thus we find that all coefficients vanish and conclude that we always have $X^{(L)\varrho}_{2L-1,N}=0$. 
%
	

	\section{All two-loop remainder functions in MRK}	
    \label{sec:results}

	In this section we apply the method described in the previous section to the symbols of all two-loop remainder functions from ref.~\cite{CaronHuot:2011ky}. As we will see, the two-loop case is special because we cannot only determine the symbols, but we can lift the symbols to the functions in a unique way. In the previous section we have seen how to extract the symbols of $X^{(2)\varrho}_{1,N}$ and $X^{(2)\varrho}_{2,N}$ and we have given analytic results for $X^{(2)\varrho}_{2L-1,N}$ and $X^{(2)\varrho}_{2L,N}$. At two loops the remainder function in MRK is then completely determined by these quantities,
		\beq\bsp
	\cM\left[R^{(2)\varrho}_N\right] &\,= (2 \pi i) X^{(2)\varrho}_{1,N} + (2 \pi i)^2 X^{(2)\varrho}_{2,N}+ (2 \pi i)^3 X^{(2)\varrho}_{3,N} + (2 \pi i)^4 X^{(2)\varrho}_{4,N}\\
	&\,= (2 \pi i) X^{(2)\varrho}_{1,N} + (2 \pi i)^2 X^{(2)\varrho}_{2,N}\,,
	\esp\eeq
	where in the last step we used $X_{3,N}^{(2)\varrho}=0$ and $X_{4,N}^{(2)\varrho}=X_{4,6}^{(2)\varrho}=0$. We thus see that at two loops the remainder function in MRK is completely determined by $X^\varrho_{1,N}$ and $X^\varrho_{2,N}$, and so we can determine their symbols using the method from the previous section. As we will discuss now, we can fix all beyond-the-symbol ambiguities at two loops and completely determine the functions $X^{(2)\varrho}_{1,N}$ and $X^{(2)\varrho}_{2,N}$.

	\subsection{The functions $X_{1,N}^{(2)\varrho}$}
	We start by extracting the symbols of $X_{1,N}^{(2)\varrho}$ from the symbol of the two-loop remainder functions in every Mandelstam region $\varrho$. This was already done in ref.~\cite{Prygarin:2011gd,Bargheer:2015djt} at LLA and NLLA for the regions where $(\varrho_3,\varrho_{N}) = (+1,+1)$, and we reproduce the results of ref.~\cite{Prygarin:2011gd,Bargheer:2015djt} for those regions. In ref.~\cite{Prygarin:2011gd,DelDuca:2018hrv} the symbols were lifted to functions in those regions. Here we follow the same reasoning to obtain the functions $X_{1,N}^{(2)\varrho}$, including for the regions where $(\varrho_3,\varrho_{N}) \neq (+1,+1)$. The final results can be written in the compact form~\cite{Bargheer:2015djt}
		\begin{align}
	\bsp 
	\label{eq:sd-relations}
	&X_{1,N}^{(2)\varrho}=\sum_{\substack{k,l\in I_{\varrho}\\ k< l}}\left( X_{1,N}^{(2)[k,l]}-X_{1,N}^{(2)[k+1,l]}-X_{1,N}^{(2)[k,l-1]}+X_{1,N}^{(2)[k+1,l-1]}\right)\ ,
	\esp
	\end{align}
	where $I_{\varrho}$ denotes the set of elements in $\varrho$ equal to $-1$ and $[k,l]$ is the Mandelstam region 
	\beq\label{eq:[k,l]}
	[k,l] = (+\ldots+\stackrel{\hbox{\small$k$}}{\hbox{$-$}}\ldots\stackrel{\hbox{\small$l$}}{\hbox{$-$}}+\ldots+)\,.
	\eeq
    Equation~\eqref{eq:sd-relations} shows that the functions $X_{1,N}^{(2)[k,l]}$ form a spanning set for all the $X_{1,N}^{(2)\varrho}$. The elements in the spanning set  are given by \cite{Bargheer:2015djt}\,,
	\begin{align}
	\bsp 
	X_{1,N}^{(2)[jk]}\,=\, \sum_{i=j}^{k-1}\left[f(\rho_i)\log(\tau_i) + \tilde{f}(\rho_i) \right] + \sum_{i=j}^{k-2} g(\rho_i,\rho_{i+1})\ .
	\esp
	\end{align}
	The functions $f$ and $\tilde{f}$ are the LLA and NLLA six-point amplitudes in MRK in the Mandelstam region $(+--+)$~\cite{Bartels:2008sc,Lipatov:2010qg,Lipatov:2010ad}
	\begin{align}
	\bsp \label{eq:6-pt_result_f}
	f(\rho_i)\,=\,&\frac{1}{2}\log \left|\frac{1-\rho_i}{\rho_i}\right|^2 \log \left| 1-\rho_i\right|^2\ ,\\
	\tilde{f}(\rho_i)\,=\,& -4 P_3(\rho_i) -\frac{1}{3} \log^2|\rho_i|^2\log|1-\rho_i|^2 + \frac{1}{3}\log^2\left|1-\rho_i\right|^2\log\frac{\left|\rho_i\right|^6}{\left|1-\rho_i\right|^4} \\
	& - \frac{1}{2}\log\left|1-\rho_i\right|^2\log\left|\frac{1-\rho_i}{\rho_i}\right|^2\log\frac{\left|\rho_i\right|^2}{\left|1-\rho_i\right|^4}\ ,
	\esp
	\end{align}
	where
	\beq
	P_n(z) =  \Re_n \sum_{k=0}^{n-1} \frac{B_k}{k!} \log^k\! |z|^2\, \text{Li}_{n-k}(z)
	\eeq
	is the single-valued generalisation of $\text{Li}_n(z)$, $B_k$ are the Bernoulli numbers and $\Re_n$ stands for real part if $n$ is odd or imaginary part otherwise. The function $g$ appears for the first time in the seven-point amplitude in the Mandelstam region $(+---+)$. It is given by~\cite{Bargheer:2015djt,DelDuca:2018hrv}
	\begin{align}
	 \nonumber
	&g(\rho_i,\rho_{i+1}) =-2 \Bigg[P_3\left(1-\frac{\rho_i}{\rho_{i+1}}\right)+P_3\left(\frac{\rho_i \left(1-\rho_{i+1}\right)}{\rho_i-\rho_{i+1}}\right)+P_3\left(\frac{\rho_i}{\rho_{i+1}}\right)+P_3\left(\frac{1-\rho_i}{\rho_{i+1}-\rho_i}\right)\\
	&-P_3\left(\rho_{i+1}\right)-P_3\left(\rho_i\right)\Bigg] -\frac{1}{6} \log \left| 1-\rho_i\right|^2 \log \left| 1-\frac{1}{\rho_{i+1}}\right|^2 \log \left| \frac{\left(1-\rho_i\right) \left(1-\rho_{i+1}\right)}{\rho_i}\right|^2\\
\nonumber	&+  \frac{1}{6} \log \left| 1-\frac{\rho_i}{\rho_{i+1}}\right|^2\log^2\left| \frac{1-\rho_i}{\rho_i \left(1-\rho_{i+1}\right)}\right|^2-\frac{1}{6} \log \left| 1-\frac{\rho_i}{\rho_{i+1}}\right|^2\log \left| 1-\rho_{i+1}\right|^2 \log \left| \rho_{i+1}\right|^2\\
	&+\frac{1}{6} \log \left| 1-\frac{\rho_i}{\rho_{i+1}}\right|^2\log \left| 1-\frac{1}{\rho_i}\right|^2 \log \left| \rho_i\right|^2\ .
	\nonumber
	\end{align}
Equation~\eqref{eq:sd-relations} reproduces the symbol of $X_{1,N}^{(2)\varrho}$, but a priori it does not determine it completely, because in general we could add multiple zeta values multiplied by single-valued polylogarithms of lower weight. In this case the only beyond-the-symbol term that we need to consider is $\zeta_3$ multiplied by an $N$-dependent rational coefficient, because terms proportional to $\zeta_2=\frac{\pi^2}{6}$ would be contained in $X_{2,N}^{(2)\varrho}$. It is then easy to check that it is not possible to add any such term without violating the behaviour of $X_{1,N}^{(2)\varrho}$ in soft limits, cf.~eq.~\eqref{eq:X_soft}.
	
	We observe that all the functions $X_{1,N}^{(2)\varrho}$ can be obtained as linear combinations of functions from the spanning set $X_{1,N}^{(2)[k,l]}$. We have checked that there are no other relations among the $X_{1,N}^{(2)\varrho}$ except for those in eq.~\eqref{eq:sd-relations}. These relations had been obtained for the first time in ref.~\cite{Bargheer:2015djt} for the subset of regions for which $(\varrho_3,\varrho_N)=(+,+)$. 
	Our results show that the same relations extend to arbitrary regions $\varrho$.

	\subsection{The functions $X_{2,N}^{(2)\varrho}$}
\begin{figure}[h]
		\centering
		\includegraphics[width=\linewidth]{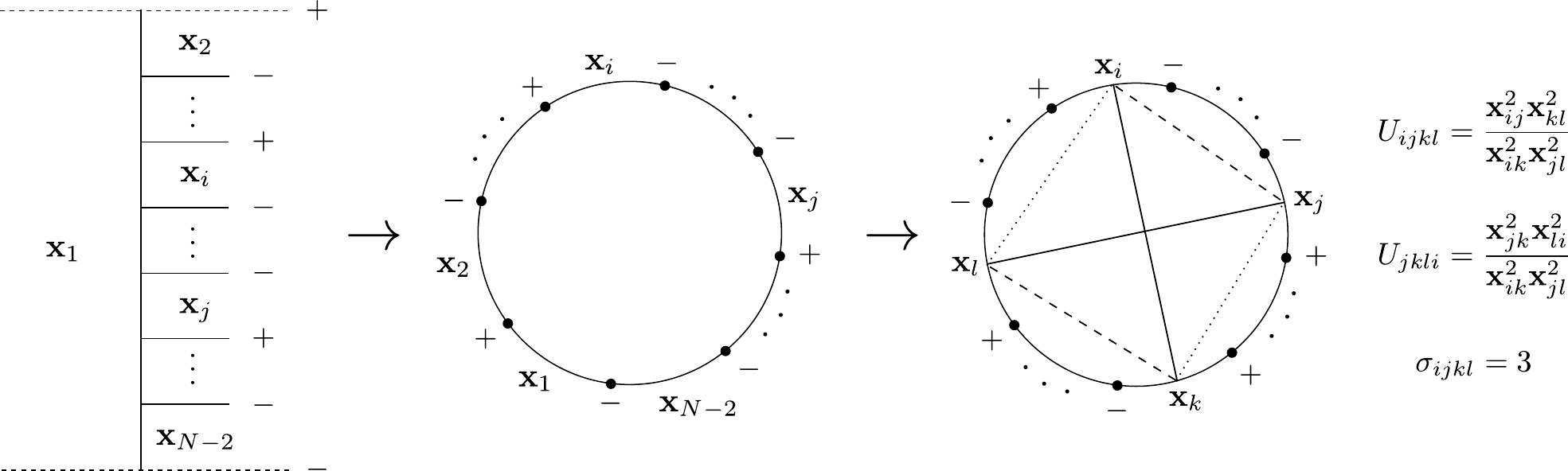}
		\caption{Illustration of the diagrammatic method for obtaining the $(2\pi i)^2$ coefficients of the remainder function, $X_{2,N}^{(2)\varrho}$.}
		\label{fig:dd-method}
	\end{figure}

	We now present the main result of this paper, the analytic results for the functions $X_{2,N}^{(2)\varrho}$. So far we have only discussed the symbols of $X_{2,N}^{(2)\varrho}$. We now argue that we can easily uplift these symbols to functions. Since $X_{2,N}^{(2)\varrho}$ is both single-valued and real, it can only be a linear combination of products of two logarithms, and it is very easy to uplift the symbols to functions. The only beyond the symbol terms that we could add is a constant proportional to $\zeta_2$, but such a constant is excluded because it would contribute to $X_{4,N}^{(2)\varrho}$ and not $X_{2,N}^{(2)\varrho}$. 
We therefore obtain the following compact result,
\begin{align}
	\label{eq:dd-general}
	X^{(2)\varrho}_{2,N}=\frac{1}{2}\sum_{C_{ijkl}}(-1)^{\sigma_{ijkl}+1} \log|U_{ijkl}|^2\log|U_{jkli}|^2\ .
	\end{align}
	The different contributions to eq.~\eqref{eq:dd-general} can be neatly obtained using a diagrammatic method described in the following and illustrated in fig.~\ref{fig:dd-method}.
	\begin{enumerate}
		\item Assemble the transverse dual variables $\bx_i$ on a circle, keeping track of the energies of the $N-2$ particles
		\item Next, draw all possible intersecting segments between four points $\bx_i,\,\bx_j,\,\bx_k,\,\bx_l$ whose adjacent particles have opposite energies. Each cross $C_{ijkl}$ defines two cross ratios.
		\item Assign to each cross a sign $(-1)^{\sigma_{ijkl}+1}$, where $\sigma_{ijkl}$ stands for the number of $+ \rightarrow -$ segments when going around the circle in the clockwise direction, ignoring all other $\bx_m$ variables. For the example in fig.~\ref{fig:dd-method}, $\sigma_{ijkl}=3$.
				  
	\end{enumerate}
	
	The functions obtained using the method above satisfy interesting properties. First of all, since the $X_{2,N}^{(2)\varrho}$ and the leading logarithmic part of $X_{1,N}^{(2)\varrho}$ (i.e., the function $f$ in eq.~\eqref{eq:6-pt_result_f}) are both products of logarithms, it is natural to ask whether there are any relations that connect the two sets of functions. Indeed, the first observation we make is that certain $X_{2,N}^{(2)\varrho}$ are identical to the leading logarithmic terms that feature in $X_{1,N}^{(2)\varrho}$, namely
		\begin{align}
		\bsp 
		\label{eq:sddd-relation}
		X_{1,N}^{(2)+\dots + - - + \dots + } \Big|_{\rm LL} = -\frac{1}{2} X_{2,N}^{(2)+\dots + - + - \dots - }\ .
		\esp
		\end{align}
	We now study the relations satisfied among the $X_{2,N}^{(2)\varrho}$ alone and arrive at a set of
    functions which constitute independent building blocks for arbitrary multiplicity. First, it
    easy to see that the $X_{2,N}^{(2)\varrho}$ are invariant under the
    simultaneous reversal of all energies,
	\begin{align}
	\label{eq:dd-relations1}
	X_{2,N}^{(2)\varrho} = X_{2,N}^{(2)\bar\varrho}\ ,
	\end{align}
	where $\bar\varrho$ denotes the region $\varrho$ with all energies reversed.
	Moreover, since the only dual variables entering the cross ratios in fig.~\ref{fig:dd-method} are bounded by particles with opposite energies, neighbouring particles with the same energy can be collapsed into one, leading to the simple rule,
	\begin{align}
	\bsp
	\label{eq:dd-relations2}
	X^{(2)\varrho_3\cdots\varrho_{i}\varrho_{i+1}\cdots\varrho_{N}}_{2,N}(\rho_1,\dots&,\rho_{N-5}) \,=\,X^{(2)\varrho_3\cdots\varrho_{i}\cdots\varrho_{N}}_{2,N-1}(\rho_1,\dots,\hat{\rho}_{i-3},\dots,\rho_{N-5}) \,,\\
	&\text{if } \varrho_i=\varrho_{i+1}\text{ and }i\notin\{3,N-1\}\ .
	\esp
	\end{align}	
	Using explicit results for amplitudes with a high number of external legs, we have searched for additional linear relations among the $X^{(2)\varrho}_{2,N}$ and the leading logarithmic part of the function $X_{1,N}^{(2)\varrho}$. We have not found any relations among these function other than eqs.~\eqref{eq:sddd-relation},~\eqref{eq:dd-relations1}~and~\eqref{eq:dd-relations2}, which leads us to conjecture that these are the only relations among these objects.

    We now present a spanning set for all the $X_{2,N}^{(2)\varrho}$. Without loss of generality, we consider regions for which particle 3 has positive energy. From the diagrammatic method above, it is clear that in order for a function $X_{2,N}^{(2)\varrho}$ to be non zero, there must be at least two blocks of consecutive particles with negative energies, e.g. $(+\cdots+-\cdots-+\cdots+-\cdots-\cdots)$.
	Moreover, our diagrammatic methods implies that regions with more than two such blocks are linear combinations of those with exactly two since, according to fig.~\ref{fig:dd-method}, they admit more than one cross. Taking into account the relations in  eq.~\eqref{eq:dd-relations2}, we find that the independent regions constitute all distinct ways of collapsing neighbouring particles of a region with two blocks of particles with negative energies. There are eight possible regions that cannot be collapsed further,
	\begin{align}
	\bsp 
	(+-+-) \quad (+-+-+)&\quad (++-+-)\quad (+-+--) \\
	(++-+-+)\quad (+-+-++)&\quad (++-+--)\quad (++-+-++)\ .
	\esp
	\end{align}	
	Therefore, we find that all the $X_{2,N}^{(2)\varrho}$ in eq.~\eqref{eq:dd-general} can be expanded in a basis of functions that arise up to nine points. Using simplicial MRK coordinates (see fig.~\ref{fig:rho}), the basis is explicitly given by
	\begin{align}
	\bsp
	\label{eq:basis} 
	F^{+-+-}(\rho_1) \,=\, & \frac{1}{2}\log \left| 1-\rho _1\right|^2  \log\left|\frac{\rho _1}{1-\rho _1}\right|^2\\[10pt]
	F^{+-+-+}(\rho_1,\rho_2) \,=\, & \frac{1}{2}\,\log\left|1-\rho_{12}\right|^2\log\left|\rho_{12}\right|^2\ ,\\[10pt]
	F^{++-+-}(\rho_1,\rho_2) \,=\, & \frac{1}{2}\,\log\left|\frac{1-\rho_1}{1-\rho_2}\right|^2\log\left|\frac{1-\rho_2}{\rho_1-\rho_2}\right|^2\ ,\\[10pt]
	F^{+-+--}(\rho_1,\rho_2) \,=\, & F^{++-+-}(1/\rho_2,1/\rho_1)\\[10pt]
	F^{++-+-+}(\rho_1,\rho_2,\rho_3) \,=\, & \frac{1}{2} \log\left| \frac{1-\rho_{13}}{1-\rho_{23}}\right|^2 \log\left| \frac{1-\rho_{21}}{1-\rho_{31}}\right|^2 \ ,\\[10pt]
	F^{+-+-++}(\rho_1,\rho_2,\rho_3) \,=\, & 	F^{++-+-+}(1/\rho_3,1/\rho_2,1/\rho_1) \ ,\\[10pt]
	F^{++-+--}(\rho_1,\rho_2,\rho_3) \,=\, & \frac{1}{2} \log\left| \frac{(1-\rho_2) (1-\rho_{31})}{(1-\rho_{21}) (1-\rho_3)}\right|^2 \log\left| \frac{(1-\rho_1) (\rho_2-\rho_3)}{(1-\rho_2) (\rho_1-\rho_3)}\right|^2 \ , \\[10pt]
	F^{++-+-++}(\rho_1,\rho_2,\rho_3,\rho_4) \,=\, & \frac{1}{2} \log\left| \frac{(1-\rho_{13}) (1-\rho_{24})}{(1-\rho_{14}) (1-\rho_{23})}\right|^2 \log\left| \frac{(1-\rho_{21})  (1-\rho_{34})}{(1-\rho_{31}) (1-\rho_{24})}\right|^2 \ ,
	\esp
	\end{align}
	with $\rho_{ij}\equiv\rho_i/\rho_j$.

\section{Discussion}
\label{sec:discussion}
In this section we analyse the explicit results for two-loop MHV remainder functions in MRK and we argue that they are consistent with the prediction from Regge theory that starting from eight points in certain Mandelstam regions the amplitudes receive contributions from an exchange of a composite state of three Reggeons~\cite{Lipatov:2009nt,Bartels:2011nz}.

Schematically, we can write
\beq\label{eq:scheme}
R_N^{\varrho}\,e^{i\pi\delta_N^\varrho} \sim 1 + i\pi\delta_N^\varrho + W_{2,N}^{\varrho} + W_{3,N}^{\varrho}+\ldots\,,
\eeq
where $R_N^{\varrho}$ denotes the all-loop MHV remainder function in the Mandelstam region $\varrho$ and $\delta_N^\varrho$ denotes a phase arising from the part of the BDS ansatz that does not undergo multi-Regge factorisation~\cite{Simon_phase},
\beq
\delta_N^{\varrho} = 
\frac{\gamma_K}{4}\sum_{i=3}^{N-3} \sum_{j=i+2}^{N-1} (\varrho_{i}-\varrho_{i+1})(\varrho_{j}-\varrho_{j+1})
  \log \left|\frac{ \bx_{ij}^2 \bx_{1 i+1} \bx_{j-1 1} } { \bx_{1 i}\bx_{i i+1} \bx_{j-1 j} \bx_{j 1} }\right|^2\,.
\eeq 
Here $\gamma_K = 4a + \mathcal{O}(a^2)$ is the cusp anomalous dimension and $a$ is the 't Hooft coupling. 
On the right-hand side, the quantities $W_{k,N}^{\varrho}$ represent schematically all the contributions involving the exchange of a composite state of exactly $k$ Reggeons. It is expected that these quantities admit a representation in terms of a dispersion integral very reminiscent of the BFKL equation in 
QCD~\cite{Lipatov:2009nt,Bartels:2011nz}. 
These dispersion integrals take the form of a (multiple) Fourier-Mellin transform, which is naturally expressed in terms of the cross ratios,
\beq\label{eq:w_def}
w_i = -\frac{(1-\rho_{i+1})(\rho_i-\rho_{i-1})}{(1-\rho_{i-1})(\rho_i-\rho_{i+1})}\,,\qquad (\rho_0,\rho_{N-4}) = (0,\infty)\,. 
\eeq
This is supported by all known results for remainder functions in MRK~\cite{Bartels:2008ce,Bartels:2008sc,Bartels:2011ge,Lipatov:2010ad,Lipatov:2010qg,DelDuca:2016lad,DelDuca:2018hrv,Bartels:2013jna,Bartels:2014jya} as well as the relation between the OPE in the near collinear limit and MRK~\cite{Basso:2014pla}.

Here we only focus on the structure of the remainder function in MRK at two loops. Projecting eq.~\eqref{eq:scheme} to two loop order, we find
\beq\label{eq:scheme-2-loops}
R_N^{(2)\varrho}+2\,(2\pi i)^2\left(\delta_N^{\varrho}\right)^2 \sim  W_{2,N}^{(2)\varrho} + W_{3,N}^{(2)\varrho}\,,
\eeq
where we have dropped terms involving an exchange of more than three Reggeons, because they are expected to contribute only at higher loop orders. Our goal is to check for small values of $N$ that our explicit results are consistent with the structure in eq.~\eqref{eq:scheme-2-loops}. In broad terms, we start from certain Mandelstam regions free of triple-Reggeon exchanges, which allows us to gain some information on which analytic structures are associated with the exchange of a composite state of two Reggeons. We then compare these analytic structures to the ones that appear in other Mandelstam regions. If two-loop MHV remainder functions were determined solely by two-Reggeon exchange, then no new analytic structures would be expected to appear in these regions. 

For simplicity, we focus on Mandelstam regions with $(\varrho_3,\varrho_N) = (+,+)$. Moreover, since it is expected that triple-Reggeon exchange only contributes to terms proportional to $(2\pi i)^2$, we only focus on that part.
More concretely, it is expected that the Mandelstam regions $[k,l]$ and $[k,l,m]$, where $[k,l]$ is defined in eq.~\eqref{eq:[k,l]} and 
\beq
	[k,l,m] = (+\ldots+\stackrel{\hbox{\small$k$}}{\hbox{$-$}}\ldots-\stackrel{\hbox{\small$l$}}{\hbox{$+$}}-\ldots\stackrel{\hbox{\small$m$}}{\hbox{$-$}}+\ldots+)\,,
	\eeq
	 do not receive contributions from three-Reggeon exchange. Our strategy is then the following:
	 \begin{enumerate}
	 \item Since the dispersion integrals are naturally expressed in terms of the variables $w_i$ rather than the $\rho_i$, we rewrite all the results of the previous section in terms of the $w_i$ variables in eq.~\eqref{eq:w_def}.
	 \item We consider the regions $[k,l]$ and $[k,l,m]$ for a fixed value of $N$ and we classify all possible products of two logarithms that appear inside $Y_N^{\varrho} \equiv X_N^{(2)\varrho} + 2\left(\delta_N^{\varrho}\right)^2$.
	 Note that $Y_N^{\varrho}$ is always a linear combination of products of two logarithms.
	 \item For any region $\varrho$ which is not of the type $[k,l]$ or $[k,l,m]$, we set to zero in $Y_N^{\varrho}$ all products of two logarithms found in the previous step. We denote the resulting function by $\cY_N^{\varrho}$. 
	 \end{enumerate}
	 
\begin{figure}[h]
		\centering
		\includegraphics[width=0.25\linewidth]{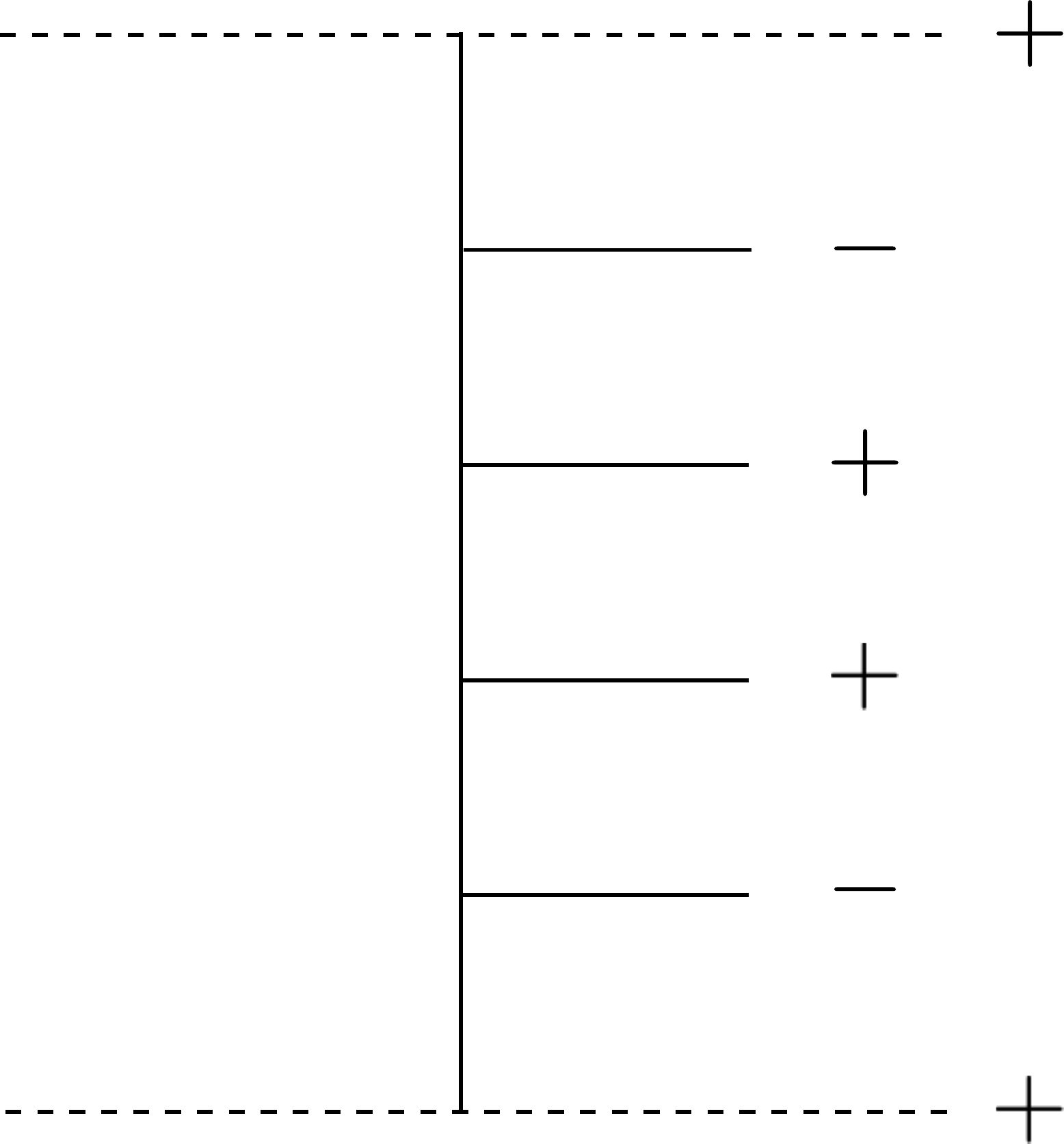}
		\caption{Region $\varrho_{8,1}$.}
		\label{fig:3reggeon8}
	\end{figure}

	 In the remainder of this section we summarise the outcome of our analysis for $N\le 9$. Let us start by discussing the case $N=8$. In that case, there is only one Mandelstam region which is not of the type $[k,l]$ or $[k,l,m]$, shown in fig.~\ref{fig:3reggeon8}, namely 
	 \beq\label{eq:varrho_{8,1}}
	 \varrho_{8,1} = (+-++-+)\,, 
	 \eeq
	 and it is expected that this region receives contributions from three-Reggeon exchange. 
	 
	 We find
	 \beq\bsp
	 \cY_8^{\varrho_{8,1}}(w_1,w_2,w_3) =&\, -3 \log \left| 1+ w_2 + w_1 w_2\right|^2 \log \left| 1+ w_3 + w_2 w_3\right|^2\\
&\,-4 \log \left| 1+ w_2\right|^2 \log \left|1 + w_3 + w_2 w_3 + w_1 w_2 w_3\right|^2\,.
\esp\eeq
We interpret the fact that we find a non-zero result as a sign of a non-trivial contribution from three-Reggeon exchange in this particular region, as expected from Regge theory.

Let us now turn to $N=9$. There are six regions that are not of the type $[k,l]$ or $[k,l,m]$, which we label as
\beq\bsp
\varrho_{9,1} &\,= (+-++-++)\,,\\
\varrho_{9,2} &\,= (++-++-+)\,,\\
\varrho_{9,3} &\,= (+-++--+)\,,\\
\varrho_{9,4} &\,= (+--++-+)\,,\\
\varrho_{9,5} &\,= (+-+++-+)\,,\\
\varrho_{9,6} &\,= (+-+-+-+)\,.
\esp\eeq
We now discuss each of these in turn. In the regions $\varrho_{9,1}$ and $\varrho_{9,2}$ the two negative-energy particles are separated by exactly two positive-energy particles, just like in eq.~\eqref{eq:varrho_{8,1}}. We therefore expect that this amplitude is identical to the corresponding eight-point amplitude (maybe up to a relabelling of the variables). This is indeed exactly what we find,
\beq\bsp
 \cY_9^{\varrho_{9,1}}(w_1,w_2,w_3,w_4) &\, =  \cY_8^{\varrho_{8,1}}(w_1,w_2,w_3)\,,\\
 \cY_9^{\varrho_{9,2}}(w_1,w_2,w_3,w_4) &\, =  \cY_8^{\varrho_{8,1}}(w_2,w_3,w_4)\,.
  \esp\eeq
For the regions $\varrho_{9,3}$ and $\varrho_{9,4}$ there is an additional negative-energy particle compared to the region $\varrho_{8,1}$. Based on experience from LLA~\cite{DelDuca:2016lad}, we expect that this amplitude reduces to an eight-point amplitude with the following relabelling of the simplicial MRK coordinates for $\varrho_{9,4}$, $(\rho_1,\rho_2,\rho_3) \to (\rho_2,\rho_3,\rho_4)$. This is indeed what we observe. More precisely, we find
\beq\bsp
 \cY_9^{\varrho_{9,3}}(w_1,w_2,w_3,w_4) &\, =  \cY_8^{\varrho_{8,1}}\left(w_1,w_2,\frac{w_3w_4}{1+w_4}\right)\,,\\
 \cY_9^{\varrho_{9,4}}(w_1,w_2,w_3,w_4) &\, =  \cY_8^{\varrho_{8,1}}\left((1+w_1)w_2,w_3,w_4\right)\,.
  \esp\eeq
  The arguments appearing in the right-hand side seem ad hoc at first, but they have a very natural interpretation and arise by expressing the cross ratios $w_i$ that appear in eight-point kinematics in terms of those in nine-point kinematics, via the intermediate of the simplicial MRK coordinates. For example, if $w_i^{(N)}$ denote the $w$-variables in $N$-point kinematics, we have in the case of the region $\varrho_{9,3}$,
  \beq\bsp
  \left(w_1^{(8)},w_2^{(8)},w_3^{(8)}\right) &\,= \left(-\frac{(1-\rho_{2})\rho_1}{(\rho_1-\rho_2)},-\frac{(1-\rho_{3})(\rho_2-\rho_{1})}{(1-\rho_{1})(\rho_2-\rho_{3})}, -\frac{\rho_3-\rho_{2}}{1-\rho_{2}}\right)\\
  &\,=\left(w_1^{(9)},w_2^{(9)},\frac{w_3^{(9)}w_4^{(9)}}{1+w_4^{(9)}}\right)\,.
  \esp\eeq
We see that our results indicate that the Mandelstam regions $\varrho_{9,i}$, $i\le 4$, receive the same three-Reggeon contributions as the eight-point region $\varrho_{8,1}$, as expected from Regge theory~\cite{Lipatov:2009nt,Bartels:2011nz}.

\begin{figure}[h]
		\centering
		\includegraphics[width=0.25\linewidth]{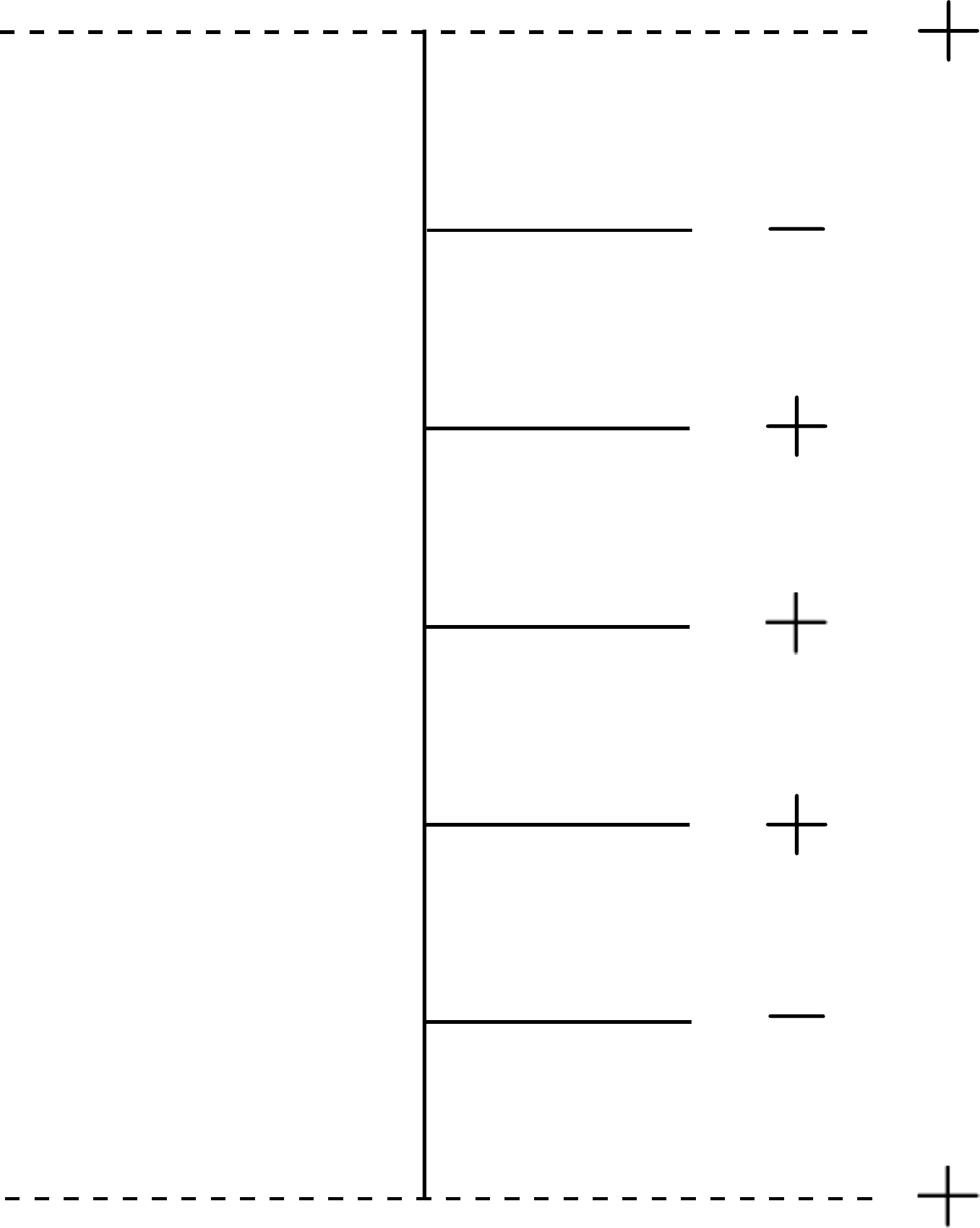}
		\caption{Region $\varrho_{9,5}$.}
		\label{fig:3reggeon9}
	\end{figure}

Let us now discuss the region $\varrho_{9,5}$, shown in fig.~\ref{fig:3reggeon9}.
	
We find that we cannot express this region in terms of an eight-point contribution only, but we have

\begin{equation}
\bsp\label{eq:Y95}
 \cY_9^{\varrho_{9,5}}&(w_1,w_2,w_3,w_4)\\
 \,= &-4 \log \left| 1+ w_3 + w_2 w_3 \right|^2 \log \left| 1+ w_4 + w_3 w_4 + w_2 w_3 w_4 + w_1 w_2 w_3 w_4\right|^2\\
& -3 \log| w_1 w_2 w_3+w_2 w_3+w_3+1|^2 \log| w_2 w_3 w_4+w_3 w_4+w_4+1|^2\,.
\esp
\end{equation}

The fact that this region contains a new analytic structure is consistent with Regge theory, because this is the first Mandelstam region sensitive to the central emission vertex for triple-Reggeon exchange. 

Finally, let us discuss the region $\varrho_{9,6}$. We find that $\cY_9^{\varrho_{9,6}}$ does not reduce to either an eight-point object nor to the function in eq.~\eqref{eq:Y95}, but instead it defines, a priori, a new analytic structure. This seems to contradict Regge theory, where no new structures are expected to appear beyond the impact factors and the central emission vertices for triple-Reggeon exchange (captured by the regions $\varrho_{8,1}$ and $\varrho_{9,5}$ respectively). Indeed, we find that $\cY_9^{\varrho_{9,6}}$ vanishes modulo the product of two logarithms encountered in $\cY_9^{\varrho_{9,i}}$, $i\le 5$, and so no new analytic structure appears in $\varrho_{9,6}$ beyond those already encountered, in agreement with the expectation.

We see that our results are fully consistent with Regge theory: there are precisely two new analytic structures that govern all Mandelstam regions at eight and nine-points, consistent with the picture of an impact factor and a central emission vertex involving a composite state of three Reggeons. 
From the BFKL perspective, new building blocks appearing in scattering amplitudes in MRK related to triple-Reggeon exchanges are expected to arise up to nine points at two loops. For the region $(+-++-+)$ at eight points, a correction to the impact factor can appear and for the region $(+-+++-+)$ at nine points the central emission block can receive extra contributions. The contributions from triple-Reggeon exchange are expected to be always accompanied by at least two powers of $(2\pi i)$, so that they contribute only to $X_{k,N}^{(2)\varrho}$ at two loops.
For future reference, we quote here the results for these two regions $(+-++-+)$ and $(+-+++-+)$, which are the simplest regions where we expect triple-Reggeon exchange to play a role. We present the results in both the $w_i$ and $\rho_i$ variables,
\beq\bsp
X^{(2)+-++-+}_{2,8}&\,= \frac{1}{2}\,\log\left|1-\frac{\rho_{1}}{\rho_3}\right|^2\log\left|\frac{\rho_{1}}{\rho_3}\right|^2
=\frac{1}{2} \log \left|\frac{w_1 w_2}{ \omega_{21}\,\omega_{32}}\right|^2 \log \left| \frac{\omega_2 \, \omega_{321}}{\omega_{21}\,  \omega_{32} }\right|^2\,,\\
X^{(2)+-+++-+}_{2,9}&\,= \frac{1}{2}\,\log\left|1-\frac{\rho_{1}}{\rho_4}\right|^2\log\left|\frac{\rho_{1}}{\rho_4}\right|^2
=\frac{1}{2} \log \left|\frac{w_1 w_2 w_3}{\omega_{321} \, \omega_{432}}\right|^2 
\log\left|\frac{\omega_{32}\, \omega_{4321}}{\omega_{321}  \, \omega_{432} }\right|^2\,,
\esp\eeq
where we introduced the shorthands
\beq
\omega_{i_1} = 1+w_{i_1} {\rm~~and~~} \omega_{i_1i_2\ldots i_m} = 1+w_{i_1}\,\omega_{i_2\ldots i_m}\,.
\eeq

\section{Conclusions}
    \label{sec:conclusions}
    In this paper we have introduced a method to extract the symbol of the coefficient of $(2\pi i)^2$ from the symbol of the remainder functions in MRK in any Mandelstam region. The cornerstone of our approach is the realisation that when the symbol is lifted to the $\Delta_{2,1,\ldots,1}$ component of the coaction on multiple polylogarithms, then the first and second entry conditions on the symbol translate into very simple polylogarithms of weight two that can appear in the first factor of the coaction. The analytic continuation of these functions is known, and so we can easily analytically continue the $\Delta_{2,1,\ldots,1}$ component of the coaction of the remainder function to any kinematic region, independently of the number of loops and the number of external legs. 
    
    We have applied our method to the symbols of all two-loop MHV amplitudes computed in ref.~\cite{CaronHuot:2011ky}, and we can uplift the symbols to functions in a unique way. The results for the coefficient of $(2\pi i)^2$ at two loops are particularly simple, and they only consist of linear combinations of products of two logarithms. We found that the two-loop results in different Mandelstam regions are not
independent but satisfy relations among each other. These relations are concisely described by
a diagrammatic method that allows us to obtain complete analytic results for all two-loop MHV remainder functions in MRK in all Mandelstam regions. In particular, we determine a spanning set for all functions that involve only building blocks with up to nine points.
    Finally, we have analysed our results and we observed that they are consistent with the hypothesis of a 
    contribution from the exchange of a three-Reggeon composite state starting from two loops and eight points in certain Mandelstam regions.
    However, we were not able to disentangle completely the contributions from two- and three-Reggeon exchange. It would be interesting to study if one can isolate the contribution from three-Reggeon exchange and to compare it to predictions by Regge theory or the OPE in the near collinear limit. We leave this for future work.

    \section*{Acknowledgements}
    We are indebted to Simon Caron-Huot for collaborating with us during the early stages of this work, and for a critical reading of the manuscript.
The authors acknowledge the support of the Higgs Center for Theoretical Physics of the University of Edinburgh in the context of the workshop ``Iterated Integrals and Multi-Regge Limit'' and of the Galileo Galilei Institute in Florence in the context of the workshop ``Amplitudes in the LHC era''. This work is supported by the ERC grant 637019 ``MathAm'' and the U.S.
Department of Energy (DOE) under contract DE-AC02-76SF00515.

	
	\appendix

	\section{MRK and Momentum Twistors}
	\label{app:momentum-twistor-MRK}

	In this appendix we set our notations and conventions. We consider scattering amplitudes of $N$ massless particles with momenta $p_i$ in MRK. It is convenient to decompose the momenta of the particles in terms of the lightcone directions and transverse coordinates,
	\beq
	p_j^{\pm} \equiv p_j^0 \pm p_j^z,\quad \pp_j\equiv p_{j\perp} =p_j^x+i p_j^y\ .
	\eeq
	In terms of these components, the scalar products between two momenta $p$ and $q$ is given by
	\beq
	2 (p \cdot q) = p^+ q^- + p^- q^+ - \pp \bar{\mathbf{q}}-\mathbf{q}\bar{\pp}\ .
	\eeq
	The mass-shell condition, $(p_j^0)^2 - (p_j^z)^2 - \pp_j\bar{\pp}_j = 0$, which in lightcone components reads $p_j^+ p_j^- - \pp_j\bar{\pp}_j = 0$,
	implies that for positive-energy states, $p_j^0 > 0$, the lightcone components are positive, $p_j^\pm > 0$, while for negative-energy states,
         $p_j^0 < 0$, they are negative, $p_j^\pm < 0$.
	
	We choose a reference frame such that the incoming particles 1 and 2 lie on the two lightcone directions,
	\beq
	p_j \equiv (p_j^+,p_j^-,\pp_j),\quad p_1 = (0,p_1^-,0),\quad 	 p_2 = (p^+_2,0,0)\ .
	\eeq
	In MRK particles $3,\ldots,N$ are strongly ordered in rapidity, which implies the hierarchy
	\beq
	|p_3^+|\gg \ldots\gg |p_N^+|,\quad 	|p_3^-|\ll \ldots\ll |p_N^-|\quad \text{and}\quad |\pp_{3}|\simeq\ldots\simeq |{\pp}_{N}|\ .
	\eeq
	Therefore, the non-trivial dynamics in the MRK regime takes place in the transverse space and particles 1 and 2 can be ignored as their transverse momenta vanish.
	
	A useful parametrisation for quantities displaying dual conformal symmetry is in terms of momentum twistors \cite{Hodges:2009hk}. Momentum twistors are obtained as a map from the spinor-helicity variables of massless momenta using the dual variables (see eq.~\eqref{eq:x}),
	\begin{align}
	\bsp
	\mu_{i\dot{\alpha}}\,&=\,\lambda_{i\alpha}\varepsilon^{\alpha\beta}x_{i\beta\dot{\alpha}}\, ,\quad \, Z_i = \begin{pmatrix}
		\lambda_{i\alpha} \\ \mu_{i\dot{\alpha}}
	\end{pmatrix}\ ,
	\esp
	\end{align}
	where
	\begin{align}
	\bsp
	\label{eq:SH}
		p_i\,&=\, x_i - x_{i-1}\,, \quad x_{ij}\equiv x_i-x_j\ , \\
	p_{\alpha\dot{\alpha}} \,&=\, \sigma_{\alpha\dot\alpha}^\mu p_{\mu} = \lambda_\alpha\tilde{\lambda}_{\dot{\alpha}}\,,\quad x_{\alpha\dot{\alpha}} = \sigma_{\alpha\dot\alpha}^\mu x_{\mu}\ .
	\esp 
	\end{align}
	In terms of the dual variables and momentum twistors, Mandelstam invariants can be written as
	\begin{align}
	(p_{i+1} + \cdots + p_{j})^2 = x_{ij}^2  = \frac{\b{i-1\,i\,j-1\,j}}{\b{i-1\, i\, I}\b{j-1\, j\, I}}\ ,
	\end{align}
	where the brackets above are defined as\footnote{In our conventions $\varepsilon^{12}=\varepsilon^{1234}=1$.}
	\begin{align}
	\bsp
	\label{eq:brackets}
	&\b{ij}\equiv \varepsilon_{\alpha\beta}\lambda_i^\alpha \lambda_j^\beta\,,\quad \alpha,\beta=1,2\ ,\\[5pt]
	&\b{i j k l} \equiv \varepsilon_{ABCD}Z_i^A Z_j^B Z_k^C Z_l^D\,,\quad A,B,C,D = 1,2,3,4\ .
	\esp
	\end{align}
	Dual conformally invariant quantities are functions of cross ratios, which in terms of momentum twistors are given by
	\begin{align}\label{eq:Us}
	U_{ijkl}\,=\,\frac{x_{ij}^2 x_{kl}^2}{x_{ik}^2 x_{jl}^2} \,=\,\frac{\b{i-1\,i\,j-1\,j}\b{k-1\,k\,l-1\,l}}{\b{i-1\,i\,k-1\,k}\b{j-1\,j\,l-1\,l}}\ .
	\end{align}
	They obey the relations
	\beq\label{eq:U_inverse}
	U_{ijkl}= U_{jilk}=U_{klij}= U_{lkji} = U_{ikjl}^{-1}= U_{jlik}^{-1} =U_{kilj}^{-1} = U_{ljki}^{-1}\ .
	\eeq
	It is also useful to note that every cross ratio $U_{ijkl}$ can be written as a product of (inverses of) cross ratios $U_{ij}$ as 
	\beq
	U_{ijkl} = \prod_{a=i}^{l-1}\prod_{b=j}^{k-1}U_{ab}^{-1}\,,\qquad i<j<k \textrm{ and } i<l\,,
	\eeq
	and the cases outside the range $i<j<k \textrm{ and } i<l$ can be worked out using eq.~\eqref{eq:U_inverse}
	
	Our strategy to compute the two-loop remainder functions in MRK is to study the $(2,1,1)$ component of their coaction.
	To this end, we need to write the first two entries of the symbols as functions of dual conformal cross ratios which can then be integrated to dilogarithms and products of logarithms, assuming the form given in eq.~\eqref{eq:211} and, in MRK, eq.~\eqref{eq:211MRL}.
	
	The first two entries of the symbol of ref.~\cite{CaronHuot:2011ky} are written in terms four-brackets of momentum twistors and combinations such as
	\begin{align}
	\b{i(ab)(cd)(ef)} \equiv \b{aicd}\b{bief}-\b{aief}\b{bicd}\ .
	\end{align}
	We replace the four-brackets involving caps in terms of cross ratios and simple brackets using\footnote{Those may be off by a sign which is irrelevant for the symbol.}:
	\begin{align}
	\begin{split}
	\b{i\,(i-1\, i+1) (j-1\,j)(j\,j+1)} &= \b{i \bar{\jmath}}\b{j\bar{\imath}} = (1-U_{ij}) \b{i-1\, i\, j-1\, j}\b{i\,i+1\,j\,j+1}\ ,
	\\[10pt]
	\b{i\,(i-1\, i+1) (j\,j+1) (k\,k+1)}& =  (1-U_{i+1\,k+1\,j+1\,i})(k+1\,i)(i+1 j+1)\ ,
	\end{split}
	\end{align}
	where in the first line we used the notation for dual twistor:
	\begin{align}
	\overline{Z_i} \equiv Z_{i-1}\wedge Z_i \wedge Z_{i+1}\ .
	\end{align}
	
	We now describe a convenient parametrisation of the multi-Regge limit of dual conformally invariant quantities in terms of momentum twistors~\cite{Prygarin:2011gd}\footnote{We fix a typo when applying the map $\mu_{i\dot{\alpha}}=\lambda_{i\alpha}\varepsilon^{\alpha\beta}x_{i\alpha\dot{\alpha}}$.}. Writing a lightlike momentum in terms of its spinor components as in eq. \eqref{eq:SH}, 
	the momentum of particle $j$ in MRK can be parametrised by:
	\begin{align}
	\begin{split}
	&p_j = \begin{pmatrix}
	\epsilon^{j-\frac{n+3}{2}} p_j^+ & \pp_j \\
	\pp_j^* &  \epsilon^{\frac{n+3}{2}-j} p_j^-
	\end{pmatrix}, \qquad j=3,\dots,n \ ,\\[10pt]
	&p_1 = -\sum_{j=3}^n \begin{pmatrix}
	0 & 0 \\
	0 &  \epsilon^{\frac{n+3}{2}-j} p_j^-
	\end{pmatrix} \ ,  \\
	&p_2 =  -\sum_{j=3}^n \begin{pmatrix}
	\epsilon^{j-\frac{n+3}{2}} p_j^+ &0 \\[10pt]
	0 & 0
	\end{pmatrix} \ ,
	\end{split}
	\end{align}
	where $\eps$ is a parameter that tends to 0 in MRK. It only serves as a bookkeeping device that can be put to unity once the limit is taken.
	The spinors associated to these momenta are then
	
	\begin{align}  
	\lambda_1=\bar{\lambda}_1 \simeq \begin{pmatrix}
	0 \\ \sqrt{p_1^-} \epsilon^{-\frac{n-3}{4}}
	\end{pmatrix}, \quad 
	&\lambda_2=\bar{\lambda}_2 \simeq \begin{pmatrix}
	\sqrt{p_2^+} \epsilon^{-\frac{n-3}{4}} \\ 0
	\end{pmatrix}\ ,\\[10pt]
	\lambda_j=\begin{pmatrix}
	\sqrt{p_j^+} \epsilon^{-\frac{n}{4} + \frac{j}{2}-\frac{3}{4}} \\ 
	\sqrt{p_j^-} \epsilon^{\frac{n}{4} - \frac{j}{2} + \frac{3}{4}} \sqrt{\frac{\pp_j}{\pp^*_j}}
	\end{pmatrix}, \quad  &\bar{\lambda}_j = \begin{pmatrix}
	\sqrt{p_j^+} \epsilon^{-\frac{n}{4} + \frac{j}{2}-\frac{3}{4}} \\ 
	\sqrt{p_j^-} \epsilon^{\frac{n}{4} - \frac{j}{2} + \frac{3}{4}} \sqrt{\frac{\pp^*_j}{\pp_j}}
	\end{pmatrix} \ .
	\end{align}
	These give the dual variables (we fix $x_n=0$)
	\begin{align}
	&x_n = \begin{pmatrix}
	0 & 0 \\ 0 & 0
	\end{pmatrix},\quad x_1 = \begin{pmatrix}
	0 & 0  \\ 0 & - \epsilon^{-\frac{n-3}{2}}\sum_{j=3}^n p_j^-\eps^{n-j}
	\end{pmatrix}\ ,\\[10pt]
	&x_j=\begin{pmatrix}
	-\epsilon^{-\frac{n-3}{2}}\sum_{k=j+1}^n p_k^+ \epsilon^{k-3} & \sum_{k=3}^{j} \pp^*_k \\
	\sum_{k=3}^{j} \pp_k & -\epsilon^{-\frac{n-3}{2}}\sum_{k=j+1}^n p_k^- \epsilon^{n-k}
	\end{pmatrix}\ .
	\end{align}
	Using  $\mu_{i\dot{\alpha}}=\lambda_{i\alpha}\varepsilon^{\alpha\beta}x_{i\alpha\dot{\alpha}}$ and $Z_i = \begin{pmatrix}
	\lambda_{i\alpha} \\ \mu_{i\dot{\alpha}}
	\end{pmatrix}$ we find:
	\begin{align}
	\begin{split}
	\label{eq:allZ}
	Z_1 = \begin{pmatrix}
	0 \\ 1 \\ 0 \\ 0
	\end{pmatrix}, \quad Z_2 = \begin{pmatrix}
	1 \\ 0 \\ 0 \\ -\sum_{k=3}^{n} \alpha_k \pp_k^* \epsilon^{\frac{n+3}{2}-k}
	\end{pmatrix}, \quad Z_n = \begin{pmatrix}
	1 \\ \alpha_n \epsilon^{-\frac{n-3}{2}} \\ 0 \\ 0
	\end{pmatrix}, \\[10pt]
	Z_j = \begin{pmatrix}
	1 \\[10pt] \epsilon^{\frac{n+3}{2}-j} \alpha_j\\[10pt]
	\sum_{k=3}^j \pp_k^* + \alpha_j \sum_{k=j+1}^n \epsilon^{k-j} \pp_k / \alpha_k \\[10pt]
	-\epsilon^{\frac{n+3}{2}-j} \alpha_j \sum_{k=3}^j  \pp_k- \sum_{k=j+1}^n \epsilon^{\frac{n+3}{2}-k}\alpha_k \pp_k^*
	\end{pmatrix}\ , \quad j=3,\dots,n-1\ ,
	\end{split}
	\end{align}
	where 
	\beq
	\alpha_j = \sqrt{\dfrac{p_j^- \pp_j}{p_j^+ \pp_j^*}} = \dfrac{\pp_j}{p^+_j}\,.
	\eeq
	Note that the last two components of $Z_j$ in eq.~\eqref{eq:allZ} are different from ref.~\cite{Prygarin:2011gd}.\\[10pt]
	We would like now to parametrise the multi-Regge limit using the variables
	\begin{equation}
	\label{eq:delta}
	\delta_i \equiv \frac{k_{i+1}^+}{k_i^+}, \quad k_i \equiv p_{i+3}\,,\quad i=0,\dots n-3 \ , \quad \delta_{n-3}\equiv 1\ . 
	\end{equation}
	Noting that 
	\begin{equation}
	\frac{\alpha_{i}}{\alpha_{i+1}} = \frac{\pp_{i}}{\pp_{i+1}}\delta_{i-3}\ ,
	\end{equation}
	we can trade $\alpha$ variables for $\delta$'s as
	\begin{align}
	\label{eq:alphadelta}
	\alpha_j = \alpha_n \frac{\pp_j}{\pp_n} \prod_{k=j-3}^{n-4} \delta_k\ .
	\end{align}
	Setting $\epsilon$ to 1 in eq.~\eqref{eq:allZ} and using eq.~\eqref{eq:alphadelta} we find:
	\begin{align}
	\begin{split}
	Z_1 = \begin{pmatrix}
	0 \\ 1 \\ 0 \\ 0
	\end{pmatrix}, \quad Z_2 = \begin{pmatrix}
	1 \\ 0 \\ 0 \\ -\frac{1}{p_n^+}\sum_{k=3}^n |\pp_k|^2 \prod_{\ell=k-3}^{n-4}\delta_\ell
	\end{pmatrix}, \quad Z_n = \begin{pmatrix}
	1 \\ \frac{\pp_n}{p_n^+} \\ 0 \\ 0
	\end{pmatrix}\ , \\
	Z_j = \begin{pmatrix}
	1  \\[10pt] \frac{\pp_j}{p_n^+} \prod_{\ell=j-3}^{n-4}\delta_\ell \\[10pt]
	\pp_j \prod_{\ell=j-3}^{n-4}\delta_\ell \sum_{k=j+1}^n \big(\prod_{\ell=k-3}^{n-4}\delta_\ell\big)^{-1} + \sum_{k=3}^j \pp_k^*\\[10pt]
	-\frac{\pp_j	}{p_n^+} \sum_{k=3}^j \pp_k \prod_{\ell=j-3}^{n-4}\delta_\ell - \frac{1}{p_n^+}\sum_{k=j+1}^{n} |\pp_k|^2 \prod_{\ell=j-3}^{n-4}\delta_\ell
	\end{pmatrix}\ .
	\end{split}
	\end{align}

\bigskip

\bibliographystyle{JHEP}
\bibliography{MRLbib}

\providecommand{\href}[2]{#2}\begingroup\raggedright\begin{thebibliography}{10}

\bibitem{Drummond:2006rz}
J.~M. Drummond, J.~Henn, V.~A. Smirnov, and E.~Sokatchev, {\it {Magic
  identities for conformal four-point integrals}},  {\em JHEP} {\bf 01} (2007)
  064, [\href{http://xxx.lanl.gov/abs/hep-th/0607160}{{\tt hep-th/0607160}}].

\bibitem{Drummond:2007au}
J.~M. Drummond, J.~Henn, G.~P. Korchemsky, and E.~Sokatchev, {\it {Conformal
  Ward identities for Wilson loops and a test of the duality with gluon
  amplitudes}},  {\em Nucl. Phys.} {\bf B826} (2010) 337--364,
  [\href{http://xxx.lanl.gov/abs/0712.1223}{{\tt arXiv:0712.1223}}].

\bibitem{Bern:2005iz}
Z.~Bern, L.~J. Dixon, and V.~A. Smirnov, {\it {Iteration of planar amplitudes
  in maximally supersymmetric Yang-Mills theory at three loops and beyond}},
  {\em Phys. Rev.} {\bf D72} (2005) 085001,
  [\href{http://xxx.lanl.gov/abs/hep-th/0505205}{{\tt hep-th/0505205}}].

\bibitem{Alday:2007hr}
L.~F. Alday and J.~M. Maldacena, {\it {Gluon scattering amplitudes at strong
  coupling}},  {\em JHEP} {\bf 06} (2007) 064,
  [\href{http://xxx.lanl.gov/abs/0705.0303}{{\tt arXiv:0705.0303}}].

\bibitem{Drummond:2007aua}
J.~M. Drummond, G.~P. Korchemsky, and E.~Sokatchev, {\it {Conformal properties
  of four-gluon planar amplitudes and Wilson loops}},  {\em Nucl. Phys.} {\bf
  B795} (2008) 385--408, [\href{http://xxx.lanl.gov/abs/0707.0243}{{\tt
  arXiv:0707.0243}}].

\bibitem{Brandhuber:2007yx}
A.~Brandhuber, P.~Heslop, and G.~Travaglini, {\it {MHV amplitudes in N=4 super
  Yang-Mills and Wilson loops}},  {\em Nucl. Phys.} {\bf B794} (2008) 231--243,
  [\href{http://xxx.lanl.gov/abs/0707.1153}{{\tt arXiv:0707.1153}}].

\bibitem{Drummond:2007cf}
J.~M. Drummond, J.~Henn, G.~P. Korchemsky, and E.~Sokatchev, {\it {On planar
  gluon amplitudes/Wilson loops duality}},  {\em Nucl. Phys.} {\bf B795} (2008)
  52--68, [\href{http://xxx.lanl.gov/abs/0709.2368}{{\tt arXiv:0709.2368}}].

\bibitem{Drummond:2008aq}
J.~M. Drummond, J.~Henn, G.~P. Korchemsky, and E.~Sokatchev, {\it {Hexagon
  Wilson loop = six-gluon MHV amplitude}},  {\em Nucl. Phys.} {\bf B815} (2009)
  142--173, [\href{http://xxx.lanl.gov/abs/0803.1466}{{\tt arXiv:0803.1466}}].

\bibitem{Alday:2007he}
L.~F. Alday and J.~Maldacena, {\it {Comments on gluon scattering amplitudes via
  AdS/CFT}},  {\em JHEP} {\bf 11} (2007) 068,
  [\href{http://xxx.lanl.gov/abs/0710.1060}{{\tt arXiv:0710.1060}}].

\bibitem{Berkovits:2008ic}
N.~Berkovits and J.~Maldacena, {\it {Fermionic T-Duality, Dual Superconformal
  Symmetry, and the Amplitude/Wilson Loop Connection}},  {\em JHEP} {\bf 09}
  (2008) 062, [\href{http://xxx.lanl.gov/abs/0807.3196}{{\tt
  arXiv:0807.3196}}].

\bibitem{CaronHuot:2010ek}
S.~Caron-Huot, {\it {Notes on the scattering amplitude / Wilson loop duality}},
   {\em JHEP} {\bf 07} (2011) 058,
  [\href{http://xxx.lanl.gov/abs/1010.1167}{{\tt arXiv:1010.1167}}].

\bibitem{Mason:2010yk}
L.~J. Mason and D.~Skinner, {\it {The Complete Planar S-matrix of N=4 SYM as a
  Wilson Loop in Twistor Space}},  {\em JHEP} {\bf 12} (2010) 018,
  [\href{http://xxx.lanl.gov/abs/1009.2225}{{\tt arXiv:1009.2225}}].

\bibitem{Alday:2010ku}
L.~F. Alday, D.~Gaiotto, J.~Maldacena, A.~Sever, and P.~Vieira, {\it {An
  Operator Product Expansion for Polygonal null Wilson Loops}},  {\em JHEP}
  {\bf 04} (2011) 088, [\href{http://xxx.lanl.gov/abs/1006.2788}{{\tt
  arXiv:1006.2788}}].

\bibitem{Gaiotto:2010fk}
D.~Gaiotto, J.~Maldacena, A.~Sever, and P.~Vieira, {\it {Bootstrapping Null
  Polygon Wilson Loops}},  {\em JHEP} {\bf 03} (2011) 092,
  [\href{http://xxx.lanl.gov/abs/1010.5009}{{\tt arXiv:1010.5009}}].

\bibitem{Gaiotto:2011dt}
D.~Gaiotto, J.~Maldacena, A.~Sever, and P.~Vieira, {\it {Pulling the straps of
  polygons}},  {\em JHEP} {\bf 12} (2011) 011,
  [\href{http://xxx.lanl.gov/abs/1102.0062}{{\tt arXiv:1102.0062}}].

\bibitem{Sever:2011da}
A.~Sever, P.~Vieira, and T.~Wang, {\it {OPE for Super Loops}},  {\em JHEP} {\bf
  11} (2011) 051, [\href{http://xxx.lanl.gov/abs/1108.1575}{{\tt
  arXiv:1108.1575}}].

\bibitem{Basso:2010in}
B.~Basso, {\it {Exciting the GKP string at any coupling}},  {\em Nucl. Phys.}
  {\bf B857} (2012) 254--334, [\href{http://xxx.lanl.gov/abs/1010.5237}{{\tt
  arXiv:1010.5237}}].

\bibitem{Basso:2013vsa}
B.~Basso, A.~Sever, and P.~Vieira, {\it {Spacetime and Flux Tube S-Matrices at
  Finite Coupling for N=4 Supersymmetric Yang-Mills Theory}},  {\em Phys. Rev.
  Lett.} {\bf 111} (2013), no.~9 091602,
  [\href{http://xxx.lanl.gov/abs/1303.1396}{{\tt arXiv:1303.1396}}].

\bibitem{Basso:2013aha}
B.~Basso, A.~Sever, and P.~Vieira, {\it {Space-time S-matrix and Flux tube
  S-matrix II. Extracting and Matching Data}},  {\em JHEP} {\bf 01} (2014) 008,
  [\href{http://xxx.lanl.gov/abs/1306.2058}{{\tt arXiv:1306.2058}}].

\bibitem{Basso:2014koa}
B.~Basso, A.~Sever, and P.~Vieira, {\it {Space-time S-matrix and Flux-tube
  S-matrix III. The two-particle contributions}},  {\em JHEP} {\bf 08} (2014)
  085, [\href{http://xxx.lanl.gov/abs/1402.3307}{{\tt arXiv:1402.3307}}].

\bibitem{Basso:2014jfa}
B.~Basso, A.~Sever, and P.~Vieira, {\it {Collinear Limit of Scattering
  Amplitudes at Strong Coupling}},  {\em Phys. Rev. Lett.} {\bf 113} (2014),
  no.~26 261604, [\href{http://xxx.lanl.gov/abs/1405.6350}{{\tt
  arXiv:1405.6350}}].

\bibitem{Basso:2014nra}
B.~Basso, A.~Sever, and P.~Vieira, {\it {Space-time S-matrix and Flux-tube
  S-matrix IV. Gluons and Fusion}},  {\em JHEP} {\bf 09} (2014) 149,
  [\href{http://xxx.lanl.gov/abs/1407.1736}{{\tt arXiv:1407.1736}}].

\bibitem{Basso:2014hfa}
B.~Basso, J.~Caetano, L.~Cordova, A.~Sever, and P.~Vieira, {\it {OPE for all
  Helicity Amplitudes}},  {\em JHEP} {\bf 08} (2015) 018,
  [\href{http://xxx.lanl.gov/abs/1412.1132}{{\tt arXiv:1412.1132}}].

\bibitem{Basso:2015rta}
B.~Basso, J.~Caetano, L.~Cordova, A.~Sever, and P.~Vieira, {\it {OPE for all
  Helicity Amplitudes II. Form Factors and Data Analysis}},  {\em JHEP} {\bf
  12} (2015) 088, [\href{http://xxx.lanl.gov/abs/1508.0298}{{\tt
  arXiv:1508.0298}}].

\bibitem{Basso:2015uxa}
B.~Basso, A.~Sever, and P.~Vieira, {\it {Hexagonal Wilson loops in planar ${
  \mathcal N }=4$ SYM theory at finite coupling}},  {\em J. Phys.} {\bf A49}
  (2016), no.~41 41LT01, [\href{http://xxx.lanl.gov/abs/1508.0304}{{\tt
  arXiv:1508.0304}}].

\bibitem{DelDuca:2009au}
V.~Del~Duca, C.~Duhr, and V.~A. Smirnov, {\it {An Analytic Result for the
  Two-Loop Hexagon Wilson Loop in N = 4 SYM}},  {\em JHEP} {\bf 03} (2010) 099,
  [\href{http://xxx.lanl.gov/abs/0911.5332}{{\tt arXiv:0911.5332}}].

\bibitem{DelDuca:2010zg}
V.~Del~Duca, C.~Duhr, and V.~A. Smirnov, {\it {The Two-Loop Hexagon Wilson Loop
  in N = 4 SYM}},  {\em JHEP} {\bf 05} (2010) 084,
  [\href{http://xxx.lanl.gov/abs/1003.1702}{{\tt arXiv:1003.1702}}].

\bibitem{Goncharov:2010jf}
A.~B. Goncharov, M.~Spradlin, C.~Vergu, and A.~Volovich, {\it {Classical
  Polylogarithms for Amplitudes and Wilson Loops}},  {\em Phys. Rev. Lett.}
  {\bf 105} (2010) 151605, [\href{http://xxx.lanl.gov/abs/1006.5703}{{\tt
  arXiv:1006.5703}}].

\bibitem{Chen:1977oja}
K.-T. Chen, {\it {Iterated path integrals}},  {\em Bull. Am. Math. Soc.} {\bf
  83} (1977) 831--879.

\bibitem{GoncharovGrassmann}
A.~B. Goncharov, {\it {A Simple Construction of the Grassmannian
  Polylogarithms}},  \href{http://xxx.lanl.gov/abs/0908.2238}{{\tt
  arXiv:0908.2238}}.

\bibitem{Brown:2009qja}
F.~C.~S. Brown, {\it {Multiple zeta values and periods of moduli spaces M 0 ,n
  ( R )}},  {\em Annales Sci. Ecole Norm. Sup.} {\bf 42} (2009) 371,
  [\href{http://xxx.lanl.gov/abs/math/0606419}{{\tt math/0606419}}].

\bibitem{Duhr:2011zq}
C.~Duhr, H.~Gangl, and J.~R. Rhodes, {\it {From polygons and symbols to
  polylogarithmic functions}},  {\em JHEP} {\bf 10} (2012) 075,
  [\href{http://xxx.lanl.gov/abs/1110.0458}{{\tt arXiv:1110.0458}}].

\bibitem{CaronHuot:2011ky}
S.~Caron-Huot, {\it {Superconformal symmetry and two-loop amplitudes in planar
  N=4 super Yang-Mills}},  {\em JHEP} {\bf 12} (2011) 066,
  [\href{http://xxx.lanl.gov/abs/1105.5606}{{\tt arXiv:1105.5606}}].

\bibitem{Golden:2013lha}
J.~Golden and M.~Spradlin, {\it {The differential of all two-loop MHV
  amplitudes in $\mathcal{N}$ = 4 Yang-Mills theory}},  {\em JHEP} {\bf 09}
  (2013) 111, [\href{http://xxx.lanl.gov/abs/1306.1833}{{\tt
  arXiv:1306.1833}}].

\bibitem{Golden:2014xqf}
J.~Golden and M.~Spradlin, {\it {An analytic result for the two-loop
  seven-point MHV amplitude in $ \mathcal{N} $ = 4 SYM}},  {\em JHEP} {\bf 08}
  (2014) 154, [\href{http://xxx.lanl.gov/abs/1406.2055}{{\tt
  arXiv:1406.2055}}].

\bibitem{Dixon:2011pw}
L.~J. Dixon, J.~M. Drummond, and J.~M. Henn, {\it {Bootstrapping the three-loop
  hexagon}},  {\em JHEP} {\bf 11} (2011) 023,
  [\href{http://xxx.lanl.gov/abs/1108.4461}{{\tt arXiv:1108.4461}}].

\bibitem{Dixon:2011nj}
L.~J. Dixon, J.~M. Drummond, and J.~M. Henn, {\it {Analytic result for the
  two-loop six-point NMHV amplitude in N=4 super Yang-Mills theory}},  {\em
  JHEP} {\bf 01} (2012) 024, [\href{http://xxx.lanl.gov/abs/1111.1704}{{\tt
  arXiv:1111.1704}}].

\bibitem{Dixon:2013eka}
L.~J. Dixon, J.~M. Drummond, M.~von Hippel, and J.~Pennington, {\it {Hexagon
  functions and the three-loop remainder function}},  {\em JHEP} {\bf 12}
  (2013) 049, [\href{http://xxx.lanl.gov/abs/1308.2276}{{\tt
  arXiv:1308.2276}}].

\bibitem{Dixon:2014iba}
L.~J. Dixon and M.~von Hippel, {\it {Bootstrapping an NMHV amplitude through
  three loops}},  {\em JHEP} {\bf 10} (2014) 065,
  [\href{http://xxx.lanl.gov/abs/1408.1505}{{\tt arXiv:1408.1505}}].

\bibitem{Dixon:2014voa}
L.~J. Dixon, J.~M. Drummond, C.~Duhr, and J.~Pennington, {\it {The four-loop
  remainder function and multi-Regge behavior at NNLLA in planar N = 4
  super-Yang-Mills theory}},  {\em JHEP} {\bf 06} (2014) 116,
  [\href{http://xxx.lanl.gov/abs/1402.3300}{{\tt arXiv:1402.3300}}].

\bibitem{Dixon:2015iva}
L.~J. Dixon, M.~von Hippel, and A.~J. McLeod, {\it {The four-loop six-gluon
  NMHV ratio function}},  {\em JHEP} {\bf 01} (2016) 053,
  [\href{http://xxx.lanl.gov/abs/1509.0812}{{\tt arXiv:1509.0812}}].

\bibitem{Caron-Huot:2016owq}
S.~Caron-Huot, L.~J. Dixon, A.~McLeod, and M.~von Hippel, {\it {Bootstrapping a
  Five-Loop Amplitude Using Steinmann Relations}},  {\em Phys. Rev. Lett.} {\bf
  117} (2016), no.~24 241601, [\href{http://xxx.lanl.gov/abs/1609.0066}{{\tt
  arXiv:1609.0066}}].

\bibitem{Drummond:2014ffa}
J.~M. Drummond, G.~Papathanasiou, and M.~Spradlin, {\it {A Symbol of
  Uniqueness: The Cluster Bootstrap for the 3-Loop MHV Heptagon}},  {\em JHEP}
  {\bf 03} (2015) 072, [\href{http://xxx.lanl.gov/abs/1412.3763}{{\tt
  arXiv:1412.3763}}].

\bibitem{Dixon:2016nkn}
L.~J. Dixon, J.~Drummond, T.~Harrington, A.~J. McLeod, G.~Papathanasiou, and
  M.~Spradlin, {\it {Heptagons from the Steinmann Cluster Bootstrap}},  {\em
  JHEP} {\bf 02} (2017) 137, [\href{http://xxx.lanl.gov/abs/1612.0897}{{\tt
  arXiv:1612.0897}}].

\bibitem{Steinmann1}
O.~Steinmann, {\it {\"Uber den Zusammenhang zwischen den Wightmanfunktionen und
  den retardierten Kommutatoren}},  {\em Acta. Phys. Hel.} {\bf 33} (1960) 257.

\bibitem{Steinmann2}
O.~Steinmann, {\it {Wightman-Funktionen und retardierte Kommutatoren II}},
  {\em Acta. Phys. Hel.} {\bf 33} (1960) 347.

\bibitem{Steinmann3}
K.~E. Cahill and H.~P. Stapp, {\it {Optical Theorems and Steinmann relations}},
   {\em Annals. Phys} {\bf 90} (1975) 438.

\bibitem{Golden:2013xva}
J.~Golden, A.~B. Goncharov, M.~Spradlin, C.~Vergu, and A.~Volovich, {\it
  {Motivic Amplitudes and Cluster Coordinates}},  {\em JHEP} {\bf 01} (2014)
  091, [\href{http://xxx.lanl.gov/abs/1305.1617}{{\tt arXiv:1305.1617}}].

\bibitem{Golden:2014xqa}
J.~Golden, M.~F. Paulos, M.~Spradlin, and A.~Volovich, {\it {Cluster
  Polylogarithms for Scattering Amplitudes}},  {\em J. Phys.} {\bf A47} (2014),
  no.~47 474005, [\href{http://xxx.lanl.gov/abs/1401.6446}{{\tt
  arXiv:1401.6446}}].

\bibitem{Golden:2014pua}
J.~Golden and M.~Spradlin, {\it {A Cluster Bootstrap for Two-Loop MHV
  Amplitudes}},  {\em JHEP} {\bf 02} (2015) 002,
  [\href{http://xxx.lanl.gov/abs/1411.3289}{{\tt arXiv:1411.3289}}].

\bibitem{Harrington:2015bdt}
T.~Harrington and M.~Spradlin, {\it {Cluster Functions and Scattering
  Amplitudes for Six and Seven Points}},  {\em JHEP} {\bf 07} (2017) 016,
  [\href{http://xxx.lanl.gov/abs/1512.0791}{{\tt arXiv:1512.0791}}].

\bibitem{Drummond:2017ssj}
J.~Drummond, J.~Foster, and {\"O}.~G{\"u}rdo{\u g}an, {\it {Cluster Adjacency
  Properties of Scattering Amplitudes in $N=4$ Supersymmetric Yang-Mills
  Theory}},  {\em Phys. Rev. Lett.} {\bf 120} (2018), no.~16 161601,
  [\href{http://xxx.lanl.gov/abs/1710.1095}{{\tt arXiv:1710.1095}}].

\bibitem{Drummond:2018dfd}
J.~Drummond, J.~Foster, and {\"O}.~G{\"u}rdo{\u g}an, {\it {Cluster adjacency
  beyond MHV}},  \href{http://xxx.lanl.gov/abs/1810.0814}{{\tt
  arXiv:1810.0814}}.

\bibitem{Golden:2018gtk}
J.~Golden and A.~J. McLeod, {\it {Cluster Algebras and the Subalgebra
  Constructibility of the Seven-Particle Remainder Function}},
  \href{http://xxx.lanl.gov/abs/1810.1218}{{\tt arXiv:1810.1218}}.

\bibitem{Bartels:2008ce}
J.~Bartels, L.~N. Lipatov, and A.~Sabio~Vera, {\it {BFKL Pomeron, Reggeized
  gluons and Bern-Dixon-Smirnov amplitudes}},  {\em Phys. Rev.} {\bf D80}
  (2009) 045002, [\href{http://xxx.lanl.gov/abs/0802.2065}{{\tt
  arXiv:0802.2065}}].

\bibitem{Bartels:2008sc}
J.~Bartels, L.~N. Lipatov, and A.~Sabio~Vera, {\it {N=4 supersymmetric Yang
  Mills scattering amplitudes at high energies: The Regge cut contribution}},
  {\em Eur. Phys. J.} {\bf C65} (2010) 587--605,
  [\href{http://xxx.lanl.gov/abs/0807.0894}{{\tt arXiv:0807.0894}}].

\bibitem{Brower:2008nm}
R.~C. Brower, H.~Nastase, H.~J. Schnitzer, and C.-I. Tan, {\it {Implications of
  multi-Regge limits for the Bern-Dixon-Smirnov conjecture}},  {\em Nucl.
  Phys.} {\bf B814} (2009) 293--326,
  [\href{http://xxx.lanl.gov/abs/0801.3891}{{\tt arXiv:0801.3891}}].

\bibitem{Brower:2008ia}
R.~C. Brower, H.~Nastase, H.~J. Schnitzer, and C.-I. Tan, {\it {Analyticity for
  Multi-Regge Limits of the Bern-Dixon-Smirnov Amplitudes}},  {\em Nucl. Phys.}
  {\bf B822} (2009) 301--347, [\href{http://xxx.lanl.gov/abs/0809.1632}{{\tt
  arXiv:0809.1632}}].

\bibitem{DelDuca:2008jg}
V.~Del~Duca, C.~Duhr, and E.~W.~N. Glover, {\it {Iterated amplitudes in the
  high-energy limit}},  {\em JHEP} {\bf 12} (2008) 097,
  [\href{http://xxx.lanl.gov/abs/0809.1822}{{\tt arXiv:0809.1822}}].

\bibitem{Kuraev:1976ge}
E.~A. Kuraev, L.~N. Lipatov, and V.~S. Fadin, {\it {Multi - Reggeon Processes
  in the Yang-Mills Theory}},  {\em Sov. Phys. JETP} {\bf 44} (1976) 443--450.
  [Zh. Eksp. Teor. Fiz.71,840(1976)].

\bibitem{Kuraev:1977fs}
E.~A. Kuraev, L.~N. Lipatov, and V.~S. Fadin, {\it {The Pomeranchuk Singularity
  in Nonabelian Gauge Theories}},  {\em Sov. Phys. JETP} {\bf 45} (1977)
  199--204. [Zh. Eksp. Teor. Fiz.72,377(1977)].

\bibitem{Balitsky:1978ic}
I.~I. Balitsky and L.~N. Lipatov, {\it {The Pomeranchuk Singularity in Quantum
  Chromodynamics}},  {\em Sov. J. Nucl. Phys.} {\bf 28} (1978) 822--829. [Yad.
  Fiz.28,1597(1978)].

\bibitem{Fadin:1998py}
V.~S. Fadin and L.~N. Lipatov, {\it {BFKL pomeron in the next-to-leading
  approximation}},  {\em Phys. Lett.} {\bf B429} (1998) 127--134,
  [\href{http://xxx.lanl.gov/abs/hep-ph/9802290}{{\tt hep-ph/9802290}}].

\bibitem{Camici:1997ij}
G.~Camici and M.~Ciafaloni, {\it {Irreducible part of the next-to-leading BFKL
  kernel}},  {\em Phys. Lett.} {\bf B412} (1997) 396--406,
  [\href{http://xxx.lanl.gov/abs/hep-ph/9707390}{{\tt hep-ph/9707390}}].
  [Erratum: Phys. Lett.B417,390(1998)].

\bibitem{Ciafaloni:1998gs}
M.~Ciafaloni and G.~Camici, {\it {Energy scale(s) and next-to-leading BFKL
  equation}},  {\em Phys. Lett.} {\bf B430} (1998) 349--354,
  [\href{http://xxx.lanl.gov/abs/hep-ph/9803389}{{\tt hep-ph/9803389}}].

\bibitem{Lipatov:2010qg}
L.~N. Lipatov and A.~Prygarin, {\it {Mandelstam cuts and light-like Wilson
  loops in N=4 SUSY}},  {\em Phys. Rev.} {\bf D83} (2011) 045020,
  [\href{http://xxx.lanl.gov/abs/1008.1016}{{\tt arXiv:1008.1016}}].

\bibitem{Bartels:2011ge}
J.~Bartels, A.~Kormilitzin, L.~N. Lipatov, and A.~Prygarin, {\it {BFKL approach
  and $2 \to 5$ maximally helicity violating amplitude in ${\cal N}=4$
  super-Yang-Mills theory}},  {\em Phys. Rev.} {\bf D86} (2012) 065026,
  [\href{http://xxx.lanl.gov/abs/1112.6366}{{\tt arXiv:1112.6366}}].

\bibitem{DelDuca:2018hrv}
V.~Del~Duca, S.~Druc, J.~Drummond, C.~Duhr, F.~Dulat, R.~Marzucca,
  G.~Papathanasiou, and B.~Verbeek, {\it {The seven-gluon amplitude in
  multi-Regge kinematics beyond leading logarithmic accuracy}},  {\em JHEP}
  {\bf 06} (2018) 116, [\href{http://xxx.lanl.gov/abs/1801.1060}{{\tt
  arXiv:1801.1060}}].

\bibitem{Basso:2014pla}
B.~Basso, S.~Caron-Huot, and A.~Sever, {\it {Adjoint BFKL at finite coupling: a
  short-cut from the collinear limit}},  {\em JHEP} {\bf 01} (2015) 027,
  [\href{http://xxx.lanl.gov/abs/1407.3766}{{\tt arXiv:1407.3766}}].

\bibitem{Dixon:2012yy}
L.~J. Dixon, C.~Duhr, and J.~Pennington, {\it {Single-valued harmonic
  polylogarithms and the multi-Regge limit}},  {\em JHEP} {\bf 10} (2012) 074,
  [\href{http://xxx.lanl.gov/abs/1207.0186}{{\tt arXiv:1207.0186}}].

\bibitem{DelDuca:2016lad}
V.~Del~Duca, S.~Druc, J.~Drummond, C.~Duhr, F.~Dulat, R.~Marzucca,
  G.~Papathanasiou, and B.~Verbeek, {\it {Multi-Regge kinematics and the moduli
  space of Riemann spheres with marked points}},  {\em JHEP} {\bf 08} (2016)
  152, [\href{http://xxx.lanl.gov/abs/1606.0880}{{\tt arXiv:1606.0880}}].

\bibitem{BrownSVHPLs}
F.~C.~S. Brown, {\it Single-valued multiple polylogarithms in one variable},
  {\em C. R. Acad. Sci. Paris, Ser. I} {\bf 338} (2004) 527.

\bibitem{BrownSVMPLs}
F.~C.~S. Brown, {\it {Single-valued hyperlogarithms and unipotent differential
  equations}},  {\em \verb+http://www.ihes.fr/~brown/RHpaper5.pdf+}.

\bibitem{Brown:2013gia}
F.~Brown, {\it {Single-valued periods and multiple zeta values}},  {\em {}}
  (2013) [\href{http://xxx.lanl.gov/abs/1309.5309}{{\tt arXiv:1309.5309}}].

\bibitem{Brown_Notes}
F.~C.~S. Brown, {\it {Notes on motivic periods}},  {\em {}} (2015)
  [\href{http://xxx.lanl.gov/abs/1512.0641}{{\tt arXiv:1512.0641}}].

\bibitem{BramRobin}
R.~Marzucca and B.~Verbeek, {\it in preparation}, .

\bibitem{Bargheer:2015djt}
T.~Bargheer, G.~Papathanasiou, and V.~Schomerus, {\it {The Two-Loop Symbol of
  all Multi-Regge Regions}},  {\em JHEP} {\bf 05} (2016) 012,
  [\href{http://xxx.lanl.gov/abs/1512.0762}{{\tt arXiv:1512.0762}}].

\bibitem{Bargheer:2016eyp}
T.~Bargheer, {\it {Systematics of the Multi-Regge Three-Loop Symbol}},  {\em
  JHEP} {\bf 11} (2017) 077, [\href{http://xxx.lanl.gov/abs/1606.0764}{{\tt
  arXiv:1606.0764}}].

\bibitem{Bartels:2014mka}
J.~Bartels, V.~Schomerus, and M.~Sprenger, {\it {The Bethe roots of Regge cuts
  in strongly coupled $ \mathcal{N}=4 $ SYM theory}},  {\em JHEP} {\bf 07}
  (2015) 098, [\href{http://xxx.lanl.gov/abs/1411.2594}{{\tt
  arXiv:1411.2594}}].

\bibitem{Bartels:2012gq}
J.~Bartels, V.~Schomerus, and M.~Sprenger, {\it {Multi-Regge Limit of the
  n-Gluon Bubble Ansatz}},  {\em JHEP} {\bf 11} (2012) 145,
  [\href{http://xxx.lanl.gov/abs/1207.4204}{{\tt arXiv:1207.4204}}].

\bibitem{Duhr:2012fh}
C.~Duhr, {\it {Hopf algebras, coproducts and symbols: an application to Higgs
  boson amplitudes}},  {\em JHEP} {\bf 08} (2012) 043,
  [\href{http://xxx.lanl.gov/abs/1203.0454}{{\tt arXiv:1203.0454}}].

\bibitem{Brown:coaction}
F.~Brown, {\it {Notes on Motivic Periods}},  {\em Commun. Num. Theor Phys.}
  {\bf 11} (2017) 557--655, [\href{http://xxx.lanl.gov/abs/1512.0641}{{\tt
  arXiv:1512.0641}}].

\bibitem{Duplancic:2000sk}
G.~Duplancic and B.~Nizic, {\it {Dimensionally regulated one loop box scalar
  integrals with massless internal lines}},  {\em Eur. Phys. J.} {\bf C20}
  (2001) 357--370, [\href{http://xxx.lanl.gov/abs/hep-ph/0006249}{{\tt
  hep-ph/0006249}}].

\bibitem{vanHameren:2005ed}
A.~van Hameren, J.~Vollinga, and S.~Weinzierl, {\it {Automated computation of
  one-loop integrals in massless theories}},  {\em Eur. Phys. J.} {\bf C41}
  (2005) 361--375, [\href{http://xxx.lanl.gov/abs/hep-ph/0502165}{{\tt
  hep-ph/0502165}}].

\bibitem{Dixon:2011ng}
L.~J. Dixon, J.~M. Drummond, and J.~M. Henn, {\it {The one-loop six-dimensional
  hexagon integral and its relation to MHV amplitudes in N=4 SYM}},  {\em JHEP}
  {\bf 06} (2011) 100, [\href{http://xxx.lanl.gov/abs/1104.2787}{{\tt
  arXiv:1104.2787}}].

\bibitem{DelDuca:2011ne}
V.~Del~Duca, C.~Duhr, and V.~A. Smirnov, {\it {The massless hexagon integral in
  D = 6 dimensions}},  {\em Phys. Lett.} {\bf B703} (2011) 363--365,
  [\href{http://xxx.lanl.gov/abs/1104.2781}{{\tt arXiv:1104.2781}}].

\bibitem{Prygarin:2011gd}
A.~Prygarin, M.~Spradlin, C.~Vergu, and A.~Volovich, {\it {All Two-Loop MHV
  Amplitudes in Multi-Regge Kinematics From Applied Symbology}},  {\em Phys.
  Rev.} {\bf D85} (2012) 085019, [\href{http://xxx.lanl.gov/abs/1112.6365}{{\tt
  arXiv:1112.6365}}].

\bibitem{Lipatov:2010ad}
L.~N. Lipatov and A.~Prygarin, {\it {BFKL approach and six-particle MHV
  amplitude in N=4 super Yang-Mills}},  {\em Phys. Rev.} {\bf D83} (2011)
  125001, [\href{http://xxx.lanl.gov/abs/1011.2673}{{\tt arXiv:1011.2673}}].

\bibitem{Lipatov:2009nt}
L.~N. Lipatov, {\it {Integrability of scattering amplitudes in N=4 SUSY}},
  {\em J. Phys.} {\bf A42} (2009) 304020,
  [\href{http://xxx.lanl.gov/abs/0902.1444}{{\tt arXiv:0902.1444}}].

\bibitem{Bartels:2011nz}
J.~Bartels, L.~N. Lipatov, and A.~Prygarin, {\it {Integrable spin chains and
  scattering amplitudes}},  {\em J. Phys.} {\bf A44} (2011) 454013,
  [\href{http://xxx.lanl.gov/abs/1104.0816}{{\tt arXiv:1104.0816}}].

\bibitem{Simon_phase}
S.~Caron-Huot, {\em {Private communication}}.
\newblock 2018.

\bibitem{Bartels:2013jna}
J.~Bartels, A.~Kormilitzin, and L.~Lipatov, {\it {Analytic structure of the
  $n=7$ scattering amplitude in $\mathcal{N}=4$ SYM theory in the multi-Regge
  kinematics: Conformal Regge pole contribution}},  {\em Phys. Rev.} {\bf D89}
  (2014), no.~6 065002, [\href{http://xxx.lanl.gov/abs/1311.2061}{{\tt
  arXiv:1311.2061}}].

\bibitem{Bartels:2014jya}
J.~Bartels, A.~Kormilitzin, and L.~N. Lipatov, {\it {Analytic structure of the
  $n=7$ scattering amplitude in $\mathcal{N}=4$ theory in multi-Regge
  kinematics: Conformal Regge cut contribution}},  {\em Phys. Rev.} {\bf D91}
  (2015), no.~4 045005, [\href{http://xxx.lanl.gov/abs/1411.2294}{{\tt
  arXiv:1411.2294}}].

\bibitem{Hodges:2009hk}
A.~Hodges, {\it {Eliminating spurious poles from gauge-theoretic amplitudes}},
  {\em JHEP} {\bf 05} (2013) 135,
  [\href{http://xxx.lanl.gov/abs/0905.1473}{{\tt arXiv:0905.1473}}].

\end{thebibliography}\endgroup
\end{document}